\documentclass{article}
\usepackage{amssymb}

\begin{document}
\author{Anatoly Konechny${\,}^{1}$\footnote{Research supported by the Director, Office of Energy 
Research, Office of High Energy and Nuclear Physics, Division of High 
Energy Physics of the U.S. Department of Energy under Contract 
DE-AC03-76SF00098 and in part by the National Science Foundation grant PHY-95-14797.} \enspace  
and  Albert Schwarz${\,}^{2}$
\footnote{Research supported in part by NSF grant DMS-9801009}
 \\
${}^{1}\,$Department of Physics, University of California Berkeley \\
and \\
Theoretical Physics Group, Mail Stop 50A-5101\\
LBNL, Berkeley, CA 94720 USA \\ 
konechny@thsrv.lbl.gov\\
\\
${}^{2}\,$Department of Mathematics, University of California Davis\\
Davis, CA 95616 USA\\
  schwarz@math.ucdavis.edu}
\title{\bf Introduction to M(atrix) theory and noncommutative geometry, Part II. }
\maketitle
\begin{abstract}
This review paper is a continuation of hep-th/0012145 and it deals  primarily 
with   noncommutative ${\mathbb R}^{d}$ spaces. We start with a discussion  of  
various algebras of smooth functions on noncommutative ${\mathbb R}^{d}$  that 
have different  asymptotic behavior at infinity. We pay particular attention to the 
differences arising when working with   nonunital algebras 
and the unitized ones obtained by adjoining the unit element.   
After introducing main objects of noncommutative geometry   over those algebras 
such as inner products, modules, connections, etc., we continue with a study of soliton and 
instanton solutions in field theories defined on these spaces. The discussion of solitons  
includes the basic facts regarding the exact soliton solutions  in the Yang-Mills-Higgs  systems as well 
as an elementary discussion of approximate solitons in scalar theories in the $\theta \to \infty$ limit. 
We concentrate on the module structure and topological numbers characterizing the solitons. 
The section on instantons contains a thorough description of noncommutative ADHM construction,
a discussion of gauge triviality conditions at infinity and the structure of a  module underlying 
the ADHM instanton solution. Although some familiarity with general ideas 
of noncommutative geometry reviewed in the first part is expected from the reader, 
this part is largely independent from the first one.    
 
\end{abstract}
\large
\tableofcontents

\newpage
\section{Introduction}
 This part of the review is devoted mostly to solitons and
instantons on  noncommutative
euclidean spaces. It is largely independent of the first part.

The main examples of noncommutative spaces considered
in the first part  were noncommutative
tori and toroidal orbifolds. The corresponding algebras of
functions are unital algebras. In the commutative case
unital algebras correspond to compact spaces; one
can say  that noncommutative space is compact if the
corresponding associative algebra has a unit element.
Our exposition was based on Connes' noncommutative geometry
and we started with compact spaces  because  main definitions
and theorems are simpler in that case.  However, the complications
arising in non-unital (= non-compact) case are not  very significant.
Every algebra can be unitized by simply  adding a unit element to it;
 using this construction we reduce all problems for
non-unital algebras
to the theory of unital algebras.

Noncommutative Euclidean space - the main object of this part- is very
simple and we do not need the full strength
of the theory developed by A. Connes and his followers. However,
we will see that unitization remains a useful technical tool
even in this simple case.  We will use it in the study
of solitons and instantons. The relation of this approach to the
standard one will be thoroughly analyzed.

\section{Noncommutative ${\mathbb R}^{d}$ spaces} 
\subsection{Quantization and deformation}
We begin with a more thorough discussion of deformation quantization.
By definition classical observables are functions on a phase space or, in mathematical terminology,
on a symplectic manifold $({\cal M}, \omega )$. In quantum mechanics observables are operators acting 
on a Hilbert space $\cal H$. A quantization procedure assigns to a function $f$ on $\cal M$ an 
operator $\hat f$ acting on $\cal H$ and depending on a quantization parameter (Planck's constant) 
$h $. It is required that in the limit $h \to 0$
\begin{equation} \label{cl_limit}
\hat f \cdot \hat g \approx \widehat {fg} \, , \quad [\hat f , \hat g] \approx -ih 
\widehat{\{ f , g \}}  
\end{equation}
where $\{. , .\}$ stand for the Poisson bracket corresponding to $\omega$.

Assume $\cal M$ is a linear symplectic space with Darboux coordinates 
$(x^{1}, \dots , x^{n}; p_{1}, \dots , p_{n})$, i.e., ${\cal M} = T^{*}{\mathbb R}^{n}$ and 
the symplectic form $\omega$ has the form 
$$
\omega = dx^{1}\wedge dp_{1} + \dots + dx^{n}\wedge dp_{n} \, .
$$ 
Then we can take ${\cal H} = L_{2}({\mathbb R}^{n})$ and introduce operators
\begin{equation}\label{Shr}
\hat x^{i} \psi(x)= x^{i}\psi(x) \, , \qquad \hat p_{i}\psi(x) = ih \partial_{i}\psi(x) \, .
\end{equation}
A construction of operator $\hat f$ corresponding to a function $f(x, p)$ is not unique. If $f$ is a polynomial 
 then it is natural to replace in it $x^{i}$, $p_{j}$ by the corresponding operators. This prescription however is 
ambiguous because  $\hat x^{i}$ and  $\hat p_{i}$ do not commute. One possibility to deal with this ambiguity is 
to symmetrize with respect to all possible orders. Then, for example, $f=(x^{1})^{2}p_{1}$ is represented by 
$$
\frac{1}{3}(\hat x^{1} \hat p_{1} \hat x^{1} + (\hat x^{1})^{2}\hat p_{1} + \hat p_{1}(\hat x^{1})^{2}) \, . 
$$
This ordering prescription is called Weyl quantization. If operator $\hat f$ is obtained from a function $f(x,p)$ 
by means of Weyl quantization then we say that $f$ is a Weyl symbol of $\hat f$. 
It is easy to verify that the Weyl symbol of an exponential operator 
$exp(i(r\cdot \hat x + s\cdot \hat p))$ is $exp(i(r\cdot  x + s\cdot  p))$. (Simply by combinatorial reasons, 
if one expands a power $(r\cdot \hat x + s\cdot \hat p)^{k}$ each ordered monomial containing a given number of $\hat x$'s 
and $\hat p$'s will appear exactly ones.)  Since the transition from symbols to operators preserves 
linear relations we can conclude from the last remark that for a given polynomial  $f(x,p)$ with a Fourier 
transform $\phi(r,s)$ 
$$
f(x, p) = \int drds\, e^{i(r\cdot  x + s\cdot  p)} \phi(r,s) 
$$
the operator corresponding to Weyl quantization rule can be written as 
$$
\hat f(x, p) = \int drds\, e^{i(r\cdot  \hat x + s\cdot  \hat p)} \phi(r,s) \, . 
$$
Using formula (\ref{Shr}) and the last formula we can compute the expression for the matrix 
elements of operator $\hat f$ via its symbol
$$
\langle x |\hat f | x' \rangle = (2\pi h)^{-n} \int dp\, e^{ip\cdot (x-x')/h} f\left( \frac{x + x'}{2},p \right)  \, . 
$$
The inverse of this formula reads 
\begin{equation} \label{symbol}
f(x, p) = \int d\xi \, \langle x -\xi/2 |\hat f | x + \xi/2 \rangle e^{ip\cdot \xi/h } \, . 
\end{equation}
This formula can be taken as a definition of the Weyl symbol for an arbitrary operator $\hat f$. 
Using the last two formulas we can compute the composition law for Weyl symbols. A symbol 
$f\ast g (x, p)$
for a composition of operators $\hat f\cdot \hat g$ with symbols $f(x, p)$, $g(x,p)$ reads 
\begin{equation} \label{Moyal1}
f\ast g (x, p) = (2\pi)^{-2n}  \int d^{n}u_{1}d^{n}u_{2} d^{n}v_{1}d^{n}v_{2} 
e^{i( u_{1}\cdot v_{1} + u_{2}\cdot v_{2})} f(x + \frac{hu_{2}}{2}, p - \frac{hu_{1}}{2}) 
g(x + v_{1}, p + v_{2}) \, . 
\end{equation}
The last formula determines a product of functions on ${\mathbb R}^{2n}$. This product is called a  star product, 
or Moyal product. It is obviously associative (because the operator multiplication  is associative). 
Multiplication formula (\ref{Moyal1}) can be generalized in the following way. Consider an antisymmetric 
$d\times d$ matrix $\theta^{ij}$. We can define a multiplication of functions on ${\mathbb R}^{d}$ by the 
formula
\begin{equation} \label{Moyal2}
f\ast g(x) = (2\pi)^{-d}\int d^{d}u d^{d}v\, f(x + \frac{1}{2}\theta u)g(x + v)e^{iuv} 
\end{equation} 
where $x$, $u$, $v$ are   $d$-dimensional vectors. We do not assume here that the matrix $\theta^{ij}$ 
is nondegenerate so that the formula also works for odd-dimensional ${\mathbb R}^{d}$ spaces. 
For the  case $d=2n$ formula (\ref{Moyal1}) corresponds to the choice of the standard matrix $\theta^{ij}$
\begin{equation} \label{theta_st} 
(\theta^{ij}_{st}) = h\cdot \left( 
\begin{array}{cc}
0& 1_{n\times n} \\
-1_{n\times n}& 0
\end{array} \right) \, .
\end{equation}

It is not hard to see that as $\theta^{ij}\to 0$ the product $f\ast g(x)$ tends to 
a pointwise multiplication. Moreover the second formula in (\ref{cl_limit}) is also satisfied for 
the Poisson bracket defined as $\{f, g\} = \partial_{i}f\theta^{ij}\partial_{j}g$.


\subsection{Algebras ${\cal S}({\mathbb R}^{d}_{\theta})$, 
${\mathbb R}_{\theta}^{d}$ and $\Gamma^{m}({\mathbb R}_{\theta}^{d})$}  \label{algebras}
To construct an algebra with multiplication rule (\ref{Moyal2}) we need to specify the allowed class of functions 
$f(x)$ on ${\mathbb R}^{d}$. For example one could consider a set of 
all polynomial functions that, as one can easily check, is closed under multiplication  (\ref{Moyal2}). 
However for the purposes of  describing physical objects concentrated in a finite domain   
we also would like to consider functions decreasing at infinity. 
 As we are interested primarily in smooth structures, say for the purposes of introducing connections etc., we 
will always assume that our functions are smooth.  Also let us note that we will consider complex valued functions. 
So that the corresponding algebras are equipped with an involution coming from the complex conjugation.

One interesting class of functions is ${\cal S}({\mathbb R}^{d})$- 
the class of Schwartz functions, i.e., infinitely differentiable functions with all derivatives tending to zero 
at infinity faster than any power. It is easy to check using (\ref{Moyal2}) that   ${\cal S}({\mathbb R}^{d})$ 
is an algebra with respect to the star product defined by  a matrix $\theta^{ij}$. We denote this algebra  
${\cal S}({\mathbb R}^{d}_{\theta})$. 
If $\theta^{ij}$ is standard, formula (\ref{symbol}) gives us an isomorphism 
between ${\cal S}({\mathbb R}^{2n}_{\theta})$ and an algebra of integral operators with matrix elements from 
${\cal S}({\mathbb R}^{2n})$. This is clear from the fact that the Fourier transform of a function from 
${\cal S}({\mathbb R}^{2n})$ is again a function from this space. Using the fact that every nondegenerate 
antisymmetric matrix $\theta^{ij}$ can be brought into  standard form by means of a linear change of coordinates,
we see that ${\cal S}({\mathbb R}^{d}_{\theta})$ is isomorphic to the algebra of integral operators with 
Schwartz class kernel.

Another interesting class is the set of smooth functions decreasing at
 infinity (i.e. functions such that their derivatives of all orders exist and tend to zero at infinity). 
These functions also  form an algebra with respect 
to the star product multiplication (\ref{Moyal2}) \cite{Rieffel}. We denote this algebra ${\mathbb R}_{\theta}^{d}$.  
If two functions $f$ and $g$ tend to zero at infinity fast enough then integral (\ref{Moyal2}) is 
absolutely convergent and the star product $f\ast g$ is well defined. If we know only that the functions $f$ and $g$ 
tend to zero at infinity we cannot say that (\ref{Moyal2}) is absolutely convergent. However, using the fact 
that the integrand in  (\ref{Moyal2})  contains a factor rapidly oscillating at infinity, we can make sense 
of this integral by means of integration by parts provided that derivatives of $f$ and $g$ also tend to zero at 
infinity.

One can consider classes of functions with  a more specified  asymptotics at infinity. Of particular interest 
are classes $\Gamma_{\rho}^{m}({\mathbb R}^{d})$  of smooth functions $f(x)$ on  ${\mathbb R}^{d}$ satisfying 
\begin{equation} \label{asymptotics}
|\partial_{\alpha}f(x)|\le C_{\alpha}(1 + | x |^{2})^{ \frac{1}{2}(m- \rho|\alpha|)} \, . 
\end{equation}
Here $\alpha= (\alpha_{1}, \dots , \alpha_{d})$ is a multiindex so that 
$\partial_{\alpha}f(x)$ stands for  $ \partial_{1}^{\alpha_{1}}\cdot \dots \cdot \partial_{d}^{\alpha_{d}}f$ and 
$|\alpha| = \alpha_{1} + \dots + \alpha_{d}$. Also $\rho$ is a real number between zero and one, not including zero,     
and $m$ is a real number, $C_{\alpha}$ are positive constants. 
The condition (\ref{asymptotics}) characterizes the behavior of $f$ and its derivatives at infinity. 
In particular it says that if for large $x$ a smooth   function $f(x)$ has an asymptotics of a polynomial 
of degree less than $m$ then it belongs to $\Gamma_{\rho}^{m}({\mathbb R}^{d})$ with $\rho = 1$.  
If $m$ is negative the function falls off at infinity and each derivative increases the power of the falloff 
by $\rho$.

One can prove that if $f\in \Gamma_{\rho}^{m_{1}}({\mathbb R}^{d})$, 
$g\in \Gamma_{\rho}^{m_{2}}({\mathbb R}^{d})$ then their star product is a function belonging to 
$\Gamma_{\rho}^{m_{1}+ m_{2}}({\mathbb R}^{d})$. Thus the set of functions 
 $\Gamma_{\rho}^{m}({\mathbb R}^{d})$ for a negative $m$ forms an algebra with respect to the star product. 
If $d=2n$ and $\theta$ is nondegenerate then 
the corresponding operators acting in Hilbert space $\cal H$ (operators having Weyl symbols 
from  $\Gamma_{\rho}^{m}({\mathbb R}^{d})$) are called pseudodifferential operators 
of the class $G_{\rho}^{m}$ (a good reference on pseudodifferential operators is  \cite{Shubin}). 
In section \ref{math} we will encounter other classes of pseudodifferential operators. Throughout the text 
we distinguish between the sets of operators and the corresponding sets of functions 
(symbols of operators). The notation for the former ones will in general  include a subscript $\theta$.

The pseudodifferential operators have a well defined action on  Schwartz class functions on 
${\mathbb R}^{d/2}$. Below we will be interested only in the case $\rho = 1$. The algebra corresponding to 
the class of functions  $\Gamma_{1}^{m}({\mathbb R}^{d})$ 
equipped with the star product (\ref{Moyal2}) will be denoted as $\Gamma^{m}({\mathbb R}^{d}_{\theta})$.
Note that the algebra ${\cal S}({\mathbb R}^{d})$ consisting of 
 Schwartz class functions can be considered as an intersection 
${\cal S}({\mathbb R}^{d}_{\theta}) = \cap_{n=1}^{\infty} \Gamma^{-n}({\mathbb R}^{d}_{\theta})$.

We will be interested in  various field theories living on on a noncommutative ${\mathbb R}^{d}$ 
space. Usually we  consider fields (e.g. Yang-Mills fields) satisfying some boundary conditions 
at infinity.  The choice of algebra that we work with is determined by the conditions at infinity we have 
in mind. It is useful to consider also along with the algebras ${\cal A} = 
 \Gamma^{m}({\mathbb R}^{d}_{\theta}), {\cal S}({\mathbb R}^{d}_{\theta}), {\mathbb R}_{\theta}^{d}$ the 
corresponding unitized algebras denoted $\tilde {\cal A}$. For every algebra $\cal A$ we construct
an algebra $\tilde {\cal A}$ by adding a unit element ${\bf 1}$. The elements of   $\tilde {\cal A}$ 
can be written in the form $a + c\cdot {\bf 1}$ where $a\in {\cal A}$, $c\in {\mathbb C}$. 
If $\cal A$ is an algebra of continuous functions on ${\mathbb R}^{d}$ tending to zero at infinity
then $\tilde{ \cal A}$ is isomorphic to the algebra of continuous functions on a $d$-dimensional sphere 
$S^{d}$. In general by unitizing the algebra of continuous functions on a locally compact space $X$ 
that tend to zero at infinity we obtain the algebra of continuous functions on its one-point compactification 
$X\sqcup \{\infty \}$. This means that the transition from $\Gamma^{m}({\mathbb R}^{d}_{\theta}), 
{\cal S}({\mathbb R}^{d}_{\theta}), {\mathbb R}_{\theta}^{d}$ to  
$\tilde \Gamma^{m}({\mathbb R}^{d}_{\theta}), 
\tilde {\cal S}({\mathbb R}^{d}_{\theta}), \tilde {\mathbb R}_{\theta}^{d}$ can be considered in some sense as a
one-point compactification of a noncommutative Euclidean space, i.e., as a transition to a topological 
(but not metric!) $d$-dimensional noncommutative sphere. 


\noindent {\bf Trace.} One can define a trace on  the algebra ${\cal S}({\mathbb R}_{\theta}^{d})$. For any 
element of  ${\cal S}({\mathbb R}_{\theta}^{d})$ the trace is given by the integral 
over ${\mathbb R}^{d}$ of the function $f(x)$ representing this  element. This definition makes sense because 
the Schwartz class functions are all integrable.      
As for the  algebras $ {\mathbb R}_{\theta}^{d}$ and $ \Gamma^{m}({\mathbb R}_{\theta}^{d})$, 
in general not every element belonging to them  is integrable. So the trace is defined only for some subset of 
elements on this algebras. 
Note however that the functions from 
 $ \Gamma^{m}({\mathbb R}_{\theta}^{d})$
, $m<-d$ are integrable and the integral can be considered as a trace on
the algebra  $ \Gamma^{m}({\mathbb R}_{\theta}^{d})$ ( the trace of
commutator vanishes).

We can formally extend the  trace defined above to the unitized algebras
$\tilde {\cal A}$ by specifying the value of 
the trace on the unit element. There is no natural way to do it and this value can be arbitrarily chosen. 

In the case of nondegenerate $\theta$ the algebras above can be considered as algebras of pseudodifferential operators 
and thus one can consider the usual operator trace on them. The connection of the operator trace and the one 
given by the integral (of the symbol) can be derived from formulas  (\ref{symbol}), (\ref{Moyal2}). 
It is given by the relation 
\begin{equation} 
\int d^{n}q\, \langle q|\hat f|q\rangle = 
 \int \frac{ d^{n}qd^{n}p}{(2\pi)^{n}} \, f(q,p)   
\end{equation}
where $n=d/2$ and $q=(q_{1}, \dots , q_{n})$, $p=(p^{1}, \dots , p^{n})$ are Darboux coordinates in which $\theta^{ij}$ 
has the standard form (\ref{theta_st}) with $h=1$.
If we are not working in Darboux coordinates then the relation between
operator trace and the trace defined as an integral of symbol with
respect to arbitrary coordinate system $x^1, ...,x^d$ contains a Jacobian
of the transition to Darboux coordinates.
We have
\begin{equation} \label{Trace} 
\hat {\rm Tr} {\hat f} =|{\rm Pfaff}(2\pi\theta )|^{-1} \int  d^{d}x  \, f(x)
\end{equation}
where $ \hat {\rm Tr}$ stands for the operator trace to be distinguished from  the  trace  
given by the integral
\begin{equation}  \label{Trace2}
{\rm Tr} f \equiv \int d^{d} x\, f(x) \, . 
\end{equation} 
This trace   has the advantage of being well defined for a  degenerate $\theta$ whereas the
 operator 
trace  blows up when $\theta$ becomes degenerate.  
 

\subsection{Projective modules and endomorphisms} \label{modules}
For any algebra $\cal A$ one can consider a (left) module ${\cal A}^{N}$ consisting of column vectors with entries 
in $\cal A$. If $\cal A$ is unital then an arbitrary element  $f\in {\cal A}^{N}$ has a unique representation  
$$
f= a_{1}\cdot e^{1} + \dots + a_{N}\cdot e^{N} 
$$
where 
$$
e^{1} = \left( \begin{array}{c} 
{\bf 1}\\
0\\
\vdots \\
0 \end{array} \right) \, , \enspace 
e^{2} = \left( \begin{array}{c} 
0\\
{\bf 1}\\
\vdots \\
0 \end{array} \right) \, , \enspace \dots  \enspace \, , 
e^{N} = \left( \begin{array}{c} 
0\\
0\\
\vdots \\
{\bf 1} \end{array} \right)
$$
and all $a_{i}\in {\cal A}$. It follows from this remark that an $\cal A$-linear map $\phi : {\cal A}^{N} \to E$ 
of ${\cal A}^{N}$ into some other module $E$ is specified by the images $\phi(e^{1}), \dots , \phi(e^{N})$. 
Moreover these images can be arbitrarily chosen. In other words the elements $e^{1}, \dots , e^{N}$ constitute  
a free system of generators of  ${\cal A}^{N}$; the module  ${\cal A}^{N}$ is thus free. 
In particular if $E$ coincides with  ${\cal A}^{N}$ then $\cal A$-linear maps are endomorphisms of  ${\cal A}^{N}$. 
Regarding   $\phi(e^{1}), \dots , \phi(e^{N})$ as an $N\times N$ matrix we obtain a correspondence between 
endomorphisms of ${\cal A}^{N}$  and  $N\times N$ matrices with entries from $\cal A$. (Every endomorphism can 
be considered as a multiplication of a column vector by such a matrix from the right.)

Let us emphasize that although the module ${\cal A}^{N}$ can be also considered for non-unital algebras $\cal A$ 
it is not a free module in that case (one cannot find a system of free generators). As earlier $N\times N$ 
matrices with entries from $\cal A$ specify endomorphisms of  ${\cal A}^{N}$ but in general not 
all endomorphisms can be represented like this. For example for the algebra ${\cal A}={\cal S}({\mathbb R}_{\theta}^{d})$ 
the endomorphisms algebra of the module ${\cal A}^{1}$ includes all smooth functions with power-like behavior 
at infinity (in particular   growing) acting for left modules by  star multiplication  from the right.

A projective module $E$ over a unital algebra $\cal A$ is defined as a direct summand in ${\cal A}^{N}$. 
We can then represent $E$ as an image of a projector $p: {\cal A}^{N}\to {\cal A}^{N}$ (an endomorphism 
obeying $p^{2}=p$, $p^{\dagger}=p$). Note that we use 
the notion of projective module only for unital algebras. In that case the notion of projective module 
corresponds to the notion of a vector bundle in the commutative geometry (see the general discussion in 
section 5.1, part I). As we discussed above for any algebra $\cal A$ one can construct a unitized 
algebra $\tilde {\cal A}$. Then every $\cal A$-module $E$ can be considered as a module over $\tilde {\cal A}$
if we represent $\bf 1$ by the identity operator in $E$. It is important to note  that this trivial 
transition from an $\cal A$-module to a $\tilde {\cal A}$-module is not just an irrelevant formality.
It leads to a change of endomorphism algebra and therefore to a change of corresponding physics 
(to a change in asymptotic behavior of the allowed fields).

Let us consider modules over algebras ${\cal A}= {\cal S}({\mathbb R}^{d}_{\theta})$ and 
$\tilde {\cal A}= \tilde {\cal S}({\mathbb R}^{d}_{\theta})$. We assume the matrix $\theta^{ij}$ 
 to be nondegenerate. (We fix our attention on this particular algebra only for definiteness. 
The modification for other algebras will be discussed later.)

As we have already noted the algebra  ${\cal S}({\mathbb R}^{d}_{\theta})$ is isomorphic to 
the algebra of  integral operators acting on functions of $d/2$ variables. It is convenient 
for the future to restrict  this action on the  (invariant) subspace of Schwartz class 
functions ${\cal S}({\mathbb R}^{d/2})$ 
because all other algebras introduced above also have a well defined action on this subspace.  
Thus the space ${\cal F} = {\cal S}({\mathbb R}^{d/2})$ can be considered as an 
${\cal S}({\mathbb R}^{d}_{\theta})$-module. We can also consider $\cal F$ as an 
$ \tilde {\cal S}({\mathbb R}^{d}_{\theta})$-module  simply representing $\bf 1$ by the identity operator 

Let us give another construction of module $\cal F$ that shows that it is a projective module over 
$ \tilde {\cal S}({\mathbb R}^{d}_{\theta})$.
  Fix an element $p\in {\cal A}\subset \tilde {\cal A}$ 
such that $p \ast p = p$. For example in the case $d=2$ it is easy to check using  (\ref{Moyal2}) 
that this equality is satisfied by the function
$$
p(x) = 2e^{-(x_{1}^{2} + x_{2}^{2})/\theta}
$$
where $\theta >0$ is the parameter of noncommutativity. It is easy to compute then 
\begin{equation}\label{normal}
\int d^{2} p(x) = 2\pi \theta \, , 
\end{equation}
that according to (\ref{Trace}) corresponds to  the operator trace of $p$ being  equal to one. 
Let us come back now to the case of an arbitrary even dimension  $d$. Choose  $p$  that satisfies 
$p\ast p = p$ and in addition  require  that 
$$
\int p \, dx =  |{\rm Pfaff}(2\pi \theta )|  
$$
that as in the above $d=2$ example means that we fix the operator trace of $p$ to be one. The conditions imposed on $p$ mean that  
 it is a  projector operator on a one-dimensional 
subspace in $\cal F$. Now to construct the corresponding (left) projective module we should consider 
elements of the form $a\cdot p$, $a\in \tilde {\cal A}$. Without loss of generality one can assume that 
$p$ has matrix elements of the form  $\langle x|\hat p|x'\rangle = \bar \alpha(x)\cdot \alpha(x')$. 
Then matrix elements of an arbitrary element in our module can be written as 
$$ 
\langle x| \hat a \hat p | x'\rangle = \bar \alpha (x) \rho(x') 
$$
where $\rho(x') \in {\cal S}({\mathbb R}^{d/2})$.    
The mapping $ap \to \rho(x')$ establishes an isomorphism between the two pictures of the module $\cal F$. 
Since the element $p$ is a projector in algebra $\tilde {\cal A}$ we conclude from the second construction 
that $\cal F$ is projective.

Note that there is also a well defined action of operators $\hat x^{j}$, $j=1, \dots , d$ 
(and polynomials thereof) on $\cal F$. In fact $\cal F$ furnishes an irreducible representation of 
the Heisenberg algebra 
\begin{equation}\label{ccr}
[\hat x^{j}, \hat x^{k}] = i\theta^{jk}\cdot {\bf 1} \, .  
\end{equation}
By Stone- von Neumann theorem there is a unique  irreducible representation of the canonical commutation 
relations in Hilbert space and one can obtain all other representations by taking direct sums of several 
copies of the irreducible one. 
Note that the Schwartz class functions ${\cal S}({\mathbb R}^{d/2})$    arises naturally as a common domain 
for operators $\hat x$ within the whole $L_{2}({\mathbb R}^{d/2})$.
By examining the proof of this theorem (that uses the canonical commutation relations in the exponentiated
Weyl form) we can argue that the corresponding representation of $\tilde {\cal S}({\mathbb R}^{d}_{\theta})$ 
is also irreducible.

There is yet another construction of the module $\cal F$ frequently used in physics literature. 
By a linear transformation the set of operators $\hat x^{i}$ satisfying (\ref{ccr}) can be brought to a set 
$a_{k}$, $a^{\dagger}_{l}$, $k, l = 1, \dots , d/2$ that satisfy 
\begin{equation} \label{a}
[a_{k}, a_{l}^{\dagger}] = \delta_{kl} \, , \quad [a_{k}, a_{l}] = [a_{k}^{\dagger}, a_{l}^{\dagger}] = 0 \, .
\end{equation}
One can think then  of  $\cal F$ as of a Fock space spanned by vectors 
\begin{equation} \label{aa}
a^{\dagger}_{k_{1}}\cdot \dots \cdot a^{\dagger}_{k_{n}}|0\rangle 
\end{equation}
where $|0\rangle$ is a Fock vacuum state, annihilated by all $a_{k}$'s. 
In the representation of $\cal F$ used above in terms of functions of $d/2$ variables the vacuum vector 
$|0\rangle$ corresponds to a Gaussian function $e^{-x\cdot x}$. Clearly this function as well as 
any vector of the form (\ref{aa}) lies in ${\cal S}({\mathbb R}^{d/2})$. Imposing the same restriction 
on infinite linear combinations of vectors (\ref{aa}) is natural from the point of view that they 
constitute  a common domain of definition for operators $a_{k}$, $a^{\dagger}_{l}$ (and polynomials of them). 
Below we will frequently refer to $\cal F$ as a Fock module.

When dealing with field theories on noncommutative ${\mathbb R}^{d}$ spaces we will be mostly 
working with endomorphisms rather than with the algebra itself. It is convenient therefore 
to have left acting endomorphisms and right modules.  From now on we will work only with right modules.
The right Fock module for which we will keep the same notation $\cal F$ consists of linear functionals 
on the original Fock module. We will adopt the ket-vector notation $\langle \cdot |$ for its elements. 
The algebra $\tilde {\cal A}$ considered as an algebra of operators  on ${\cal S}({\mathbb R}^{d/2})$ 
then acts naturally as $\langle \cdot | \mapsto \langle \cdot |\hat a$, $\hat a \in \tilde {\cal A}$. 
Due to the irreducibility of the module $\cal F$ its endomorphisms are trivial. More generally 
for $k$ copies of $\cal F$ endomorphisms are $k\times k$ matrices with numerical entries, that act naturally 
from the left.

A (right) free module  of rank $N$ consists of $N$-columns of 
elements from  $\tilde {\cal A}$ and the algebra acts by multiplication from the right.
Endomorphisms of this module are $N\times N$ matrices with $\tilde {\cal A}$-valued entries, that 
act on the $N$-columns from the left in the natural way.

One can prove \cite{Connes_pr} that any projective module over $\tilde {\cal A}$ is isomorphic 
to a module ${\cal F}^{k}\oplus (\tilde {\cal A})^{N}$. Thus it is characterized 
by two nonnegative integers $k, N$. In the case of $d=2$ these integers can be thought of as a magnetic flux number and 
a rank of the gauge group, and for $d=4$ $k$ plays the role of Pontryagin number and $N$ of the rank of the gauge 
group. The $K_{0}$-group of $\tilde {\cal A}$ is therefore ${\mathbb Z}\oplus {\mathbb Z}$ that 
matches with the $K_{0}$-group of the even-dimensional sphere. The positive cone however is different. 
For instance one cannot realize modules with a negative magnetic flux or Pontryagin number. 
One can attribute 
this to a labeling problem - there is no natural way to distinguish the deformations corresponding to 
$\theta^{ij}$ and $-\theta^{ij}$.

One can represent an endomorphism $X$ of a module ${\cal F}^{k}\oplus (\tilde {\cal A})^{N}$ in 
a block form 
$$
X = \left( 
\begin{array}{cc}
A & B \\
C & D
\end{array} \right) 
$$
where $A$ is a $k\times k$ matrix with $\mathbb C$-valued entries, 
$B$ is a  $k\times N$ matrix whose entries are elements  from  $\cal F$, $C$ is a $N\times k$ matrix with 
entries from the dual space $\cal F^{*}$ and $D$ 
is an 
$N\times N$ matrix  with entries 
belonging to  
$\tilde {\cal A}$.
(Later using the fact that the algebra $\tilde {\cal A}$ acts on $\cal F$
we shall consider its elements as
operators ${\cal F} \to {\cal
F}$ .)
  In the case when the endomorphism $X$ is
hermitian
 the matrix $A$ is hermitian, $D$ is a hermitian operator, and 
$B = C^{\dagger}$. It is convenient to represent such an endomorphism in the form 
\begin{equation} \label{endomorphism}
X = \left( 
\begin{array}{cc}
A^{j}_{i} &  \langle b_{i}^{\beta} | \\
 | b^{j}_{\alpha}\rangle & \hat D_{\alpha}^{ \beta}
\end{array} \right) 
\end{equation}
where the indices $i, j$ run from $1$ to $k$ and the indices $\alpha, \beta$ run from $1$ to $N$. 
This endomorphism maps  an element 
$$
V = \left(
\begin{array}{c}
\langle v_{i}| \\
 \hat V_{\alpha}  
\end{array} \right) \in {\cal F}^{k}\oplus (\tilde {\cal A})^{N}
$$
into 
$$
\left( 
\begin{array}{cc}
A^{j}_{i} &  \langle  b_{i}^{\beta} |  \\
 | b^{j}_{\alpha}\rangle & \hat D_{\alpha}^{ \beta}
\end{array} \right) 
\left(
\begin{array}{c}
\langle v_{i}| \\
 \hat V_{\alpha}  
\end{array} \right) = 
\left(
\begin{array}{c}
 A_{i}^{j} \langle v_{j}| +  \langle b_{i}^{\beta}| \hat V_{\beta} \\
 |b^{j}_{\alpha} \rangle \langle v_{j} | + 
\hat D_{\alpha}^{\beta}\cdot \hat V_{\beta} 
\end{array} \right) \, .
 $$

The composition of two endomorphisms is also quite  transparent in  the notations (\ref{endomorphism}): 
\begin{equation}
\left( 
\begin{array}{cc}
A^{j}_{i} &  \langle b_{i}^{\beta}| \\
 | b^{j}_{\alpha}\rangle & \hat D_{\alpha}^{ \beta}
\end{array} \right) 
\left( 
\begin{array}{cc} 
 B^{k}_{j} &  \langle c_{j}^{\gamma} | \\
 | c^{k}_{\beta}\rangle & \hat E_{\beta}^{ \gamma}
\end{array} \right) = 
\left( 
\begin{array}{cc} 
 (AB)^{k}_{i} +  \langle b_{i}^{\beta}  | c^{k}_{\beta} \rangle & 
A_{i}^{j}\langle c_{j}^{\gamma}| + \langle b_{i}^{\beta}|\hat E^{\gamma}_{\beta}\\
|b_{\alpha}^{j}\rangle B_{j}^{k} + \hat D_{\alpha}^{\beta}|c_{\beta}^{k}\rangle & 
|b_{\alpha}^{j}\rangle\langle c_{j}^{\gamma}| + \hat D_{\alpha}^{\beta}\hat E_{\beta}^{\gamma}  
\end{array} \right)   
\end{equation}
where the summation over repeated upper and lower indices is assumed. 
To summarize manipulations with endomorphisms follow a simple rule of matrix multiplication plus 
writing one term next to another in the 
given order and making sense of the resulting expression as whatever it appears to look like, e.g. 
a ket-vector followed by a bra-vector is an operator (element of the algebra), same elements in the opposite 
order give rise to a number, etc.   

A normalized trace specified on  a unital algebra $\tilde {\cal A} $ gives rise to a normalized trace on an algebra of endomorphisms 
of any projective module over this algebra. If the projector module at hand is specified by 
a projection $P:\tilde {\cal A}^{N} \to\tilde {\cal A}^{N} $ then endomorphisms of the module specified by the image 
of $P$ can be considered as  a subalgebra of the matrix algebra $Mat_{N}  \tilde {\cal A}$ consisting 
of the elements of the form $PeP$, $e\in Mat_{N}  \tilde {\cal A}$. Restricting the natural trace on the matrix 
algebra to this subspace we obtain a trace on the algebra of endomorphisms.
It is not hard to derive from (\ref{Trace}) that the trace specified 
by  integration (\ref{Trace2}) induces a trace on the endomorphisms so that for a general 
endomorphism (\ref{endomorphism}) we have 
\begin{equation} \label{Trace3}
{\rm Tr}X = |{\rm Pfaff}(2\pi \theta )|  \sum_{i=1}^{k} A_{i}^{i} + 
\sum_{\alpha=1}^{N}\int d^{d}x\, D_{\alpha}^{\alpha}(x)  \, ,  
\end{equation}
where $D_{\alpha}^{\beta}(x)$ is the matrix-valued function on ${\mathbb R}^{d}$ representing the operator  
$\hat D_{\alpha}^{\beta}$.
Again it should be noted that this trace is well defined only for a certain subclass of  endomorphisms. 
  Namely the trace (\ref{Trace3}) exists when   
$D_{\alpha}^{\alpha}(x)$  is integrable. For $N> 0$ that condition evidently excludes the identity endomorphism. 
However for a module ${\cal F}^{k}$ we have 
$$
{\rm Tr}1 =  k\cdot |{\rm Pfaff}(2\pi \theta )|  = {\rm Tr} P_{k}
$$ 
where $ P_{k}$ stands for a projector operator $P_{k}: \tilde {\cal A}^{k} \to  \tilde {\cal A}^{k}$ that 
singles out    ${\cal F}^{k}$ in the corresponding free module.


\subsection{Inner products} \label{innpr}

 In section 5.4 of part I where we concentrated on  unital algebras we explained that if an algebra $\cal A$
is equipped with involution then one can define a  $\cal A$-valued inner product $<.,.>_{\cal A}$ on a free module 
${\cal A}^{N}$ by the formula 
\begin{equation}\label{inner_pr}
<e_{1},e_{2}>_{\cal A}\equiv <(a_{1}, \dots , a_{N}), (b_{1}, \dots , b_{N})>_{\cal A}=\sum_{i=1}^{N}a_{i}^{*}b_{i} \, . 
\end{equation}
Similarly we  can define an $\cal A$-valued inner product on modules ${\cal A}^{N}$ over nonunital 
algebras by exactly same formula. The inner product (\ref{inner_pr}) satisfies  
\begin{equation} \label{property}
<e_{1}a, e_{2}b>_{\cal A} = a^{*}<e_{1}, e_{2}>_{\cal A}b
\end{equation}
for any two elements $a,b\in {\cal A}$ and any two vectors $e_{1}, e_{2} \in {\cal A}^{N}$. 
In general we say that an ${\cal A}$-module $E$ is a Hilbert module if it is equipped with an ${\cal A}$-valued 
inner product satisfying (\ref{property}).
It is easy to construct the appropriate inner products on modules  ${\cal A}^{N}$ and ${\cal F}^{k}$ 
described in section \ref{modules}  over any of the algebras introduced in section \ref{algebras}.
For the first type of modules formula (\ref{inner_pr}) works. For  ${\cal F}^{k}$ the construction is as follows. 
Let $\phi, \chi \in {\cal F}^{k}$. The inner product  $< \phi , \chi >_{\cal A}$ is given by the element of $\cal A$
specified by the composition $\phi \circ \chi^{*} \in Hom({\cal F}^{k}, {\cal F}^{k})$. Using the  Dirac bra and ket 
notations the construction reads 
$$
< \phi , \chi >_{\cal A} = |\phi\rangle\langle \chi| \, . 
$$

We say that an endomorphism $\Phi: E \to E$ is hermitian if for a given $\cal A$-valued inner product $<.,.>_{\cal A}$ 
it satisfies 
$$
<e_{1}, \Phi e_{2}>_{\cal A} = <\Phi e_{1},  e_{2}>_{\cal A}
$$
for any two elements $e_{1}, e_{2}\in E$.  An endomorphism  is called antihermitian if it satisfies the above equality with 
the minus sign. For the modules  ${\cal A}^{N}\oplus {\cal F}^{k}$ discussed above this general definition is equivalent 
to the requirement that    hermitian endomorphisms should have the form   (\ref{endomorphism}). 


\subsection{Connections} \label{conn}

Translations in space ${\mathbb R}^{d}$ induce an action  of the corresponding abelian Lie group 
on the algebras $\cal A$, $\tilde {\cal A}$ where $\cal A$ is one of the algebras introduced 
in section \ref{algebras}. One defines a connection $\nabla_{i}$ on  
projective module $E$ over  $\tilde {\cal A}$ in standard way. 
Namely  $\nabla_{i}$ are  operators $E\to E$ satisfying the Leibniz rule
\begin{equation} \label{Leib}
\nabla_{i} (e \cdot f) = e\cdot (\partial_{i} f) + (\nabla_{i}(e))\cdot f 
\end{equation}
for  arbitrary $e\in E$ and $f\in \tilde {\cal A}$. Here $\partial_{i}$ stand 
for the infinitesimal action of the translation in the $i$-th direction that is represented on functions 
by the partial derivative denoted the same way. 
On a free module  $(\tilde {\cal A})^{N}$ an arbitrary connection can be written as 
$$
\nabla_{i} = \partial_{i} + (\hat A_{i})_{\alpha}^{\beta} 
$$
where $(\hat A_{i})_{\alpha}^{\beta} $ represents an endomorphism.

Consider now a module $\cal F$. As we discussed above we have a well defined action of operators $\hat X^{i}$ 
on this module. 
One can check that operators of right multiplication by $i\theta_{jk}^{-1}\hat x^{k}$ 
denoted $\nabla_{j}^{(0)}$ satisfy  the Leibniz rule (\ref{Leib}). This connection has the curvature 
$$ 
F_{jk} =  [\nabla_{j}^{(0)}, \nabla_{k}^{(0)}] = - i \theta^{-1}_{jk}  {\bf 1} \, . 
$$ 
We will work with antihermitian connections. This means that an arbitrary connection 
can be represented as $\nabla_{j} = \nabla^{0}_{j} + i A_{j}$ where $ \nabla^{0}_{j}$ is some fiducial 
(antihermitian) connection and $A_{j}$ is a hermitian endomorphism. 
We can choose a fiducial connection on a general module $E= {\cal F}^{k}\oplus (\tilde {\cal A})^{N}$ as 
\begin{equation}\label{st_conn}
\nabla^{0}_{j} = 
\left( 
\begin{array}{cc}  i\theta_{jk}^{-1}\hat x^{k} & 0\\
0& \partial_{j} \end{array} \right)
\end{equation}
This connection acts on an arbitrary element $V\in E$ as 
$$
\nabla^{0}_{j} \left(
\begin{array}{c}
\langle v| \\
 \hat V  
\end{array} \right)
= 
i(\theta^{-1})_{jk} \left(
\begin{array}{c} 
 \langle v|\hat x^{k} \\
- (\hat x^{k}\hat V - \hat V_{\alpha}\hat x^{k} )  
\end{array} \right)
$$
where for clarity we dropped the vector indices at $\langle v|$ and $V$.

An arbitrary antihermitian connection on  $E$ 
has the form 
\begin{equation} \label{connection}
\nabla_{j} = \nabla^{0}_{j} + i \left( 
\begin{array}{cc}
(A_{j})^{k}_{i} &  \langle (b_{j})_{i}^{\beta} | \\
 | (b_{j})^{k}_{\alpha}\rangle &  (\hat D_{j})_{\alpha}^{ \beta}
\end{array} \right) \, . 
\end{equation}


\noindent {\bf Complex coordinates.}  
It is often convenient to use standard  set of complex coordinates $z_{\alpha}$ 
$\alpha = 1, \dots , n$ where $n=d/2$  constructed as follows. 
Let us first bring the matrix $\theta^{ij}$ by means of an $SO(d)$ transformation  to standard form 
\begin{equation} \label{standard_theta}
\theta^{ij} = \left( 
\begin{array}{cc} 
0 & {\rm diag}(\theta_{1}, \dots , \theta_{n}) \\
- {\rm diag}(\theta_{1}, \dots , \theta_{n}) & 0 \\
\end{array}
\right)
\end{equation} 
written here in  $n\times n$ block form.

Let us introduce complex coordinates
\begin{equation} \label{cc}
z_{1} = \frac{1}{\sqrt{2}}( x^{1} + ix^{n+1})   \, ,  \frac{1}{\sqrt{2}}( x^{2} + ix^{n+2}) \, , \dots \, , 
z_{n} = \frac{1}{\sqrt{2}}( x^{n} + ix^{2n}) \, . 
\end{equation}
It is further convenient to introduce annihilation operators 
\begin{eqnarray*}
&& a_{1}=\frac{1}{\sqrt{2|\theta_{1}|}}(\hat x^{1} + i\cdot {\rm sgn}(\theta_{1})\hat x^{n +1})\, , 
a_{2}=\frac{1}{\sqrt{2|\theta_{2}|}}(\hat x^{2} + i\cdot {\rm sgn}(\theta_{2})\hat x^{n +2})\, , \\
&& \dots \, , a_{d}=\frac{1}{\sqrt{2|\theta_{n}|}}(\hat x^{n} + i\cdot {\rm sgn}(\theta_{n})\hat x^{2n})
\end{eqnarray*} 
and creation operators $a^{\dagger}_{\alpha}$ which are hermitian conjugates of $a_{\alpha}$. 
They satisfy  the canonical commutation relations (\ref{a}). The sign factors ${\rm sgn}(\theta_{\alpha})$ 
allow us to treat arbitrary $\theta^{jk}$ of the form (\ref{standard_theta}) that might have positive as well 
as  negative eigenvalues $\theta_{i}$. The correspondence between the operators $\hat z_{\alpha}$ and  $a_{\alpha}$, 
 $a^{\dagger}_{\alpha}$ now has the following form 
\begin{equation} \label{z_vs_a}
\hat z_{\alpha} =  \left\{  \begin{array}{c} 
a_{\alpha}(\theta_{\alpha})^{1/2} \, , \enspace \theta_{\alpha}>0 \\
a_{\alpha}^{\dagger}|\theta_{\alpha}|^{1/2} \, , \enspace \theta_{\alpha}<0
\end{array} \right.
\end{equation}
with the conjugated expressions for $\widehat{\bar z_{\alpha}}$. 

In coordinates $z_{\alpha}$, $\bar z_{\alpha}$ we obtain  
\begin{equation} \label{complex_conn}
\nabla_{z_{\alpha}}^{0} \left(
\begin{array}{c}
\langle v| \\
 \hat V  
\end{array} \right)
= 
\frac{1}{\theta_{\alpha}} \left(
\begin{array}{c} 
  \langle v| \widehat{\bar z}_{\alpha}  \\
-\lbrack  \widehat{\bar z}_{\alpha}, \hat V \rbrack   
\end{array} \right) \, . 
\end{equation}

 An arbitrary connection has the form 
$$
\nabla_{z_{j}} = \nabla^{0}_{z_{j}}+ \left( 
\begin{array}{cc}
A_{\alpha} &  \langle b_{\alpha} | \\
 | c_{\alpha}\rangle &  \hat D_{\alpha}
\end{array} \right) \, , \quad 
\nabla_{\bar z_{\alpha}} = (\nabla^{0}_{z_{j}})^{\dagger}  +
  \left( 
\begin{array}{cc}
A_{\alpha}^{\dagger} &  \langle c_{\alpha} | \\
 | b_{\alpha}\rangle &  \hat D_{\alpha}^{\dagger}
\end{array} \right)
$$
where for brevity we omitted the matrix indices.

\subsection{Yang-Mills and scalar fields}
Once we described endomorphisms and connections on modules over $\cal A$ and $\tilde {\cal A}$ we are ready to define field 
theories on these modules. To write  action functionals specifying these theories we first 
choose a metric tensor $g_{ij}$ on ${\mathbb R}^{d}$ with a help of which we will be raising and lowering the indices.

A connection on a module $E$ is called  a Yang-Mills field, an endomorphism is a scalar field that in some  
sense is an analogue of a (commutative) scalar field in the adjoint representation of the gauge group. 
It is also possible to define scalar fields ``in the fundamental representation'' that are elements 
of the module itself. To obtain a number-valued functional of  these fields we will be using the trace defined according 
to (\ref{Trace2}), (\ref{Trace3}). 

Let us first explain  the case when the scalar fields are endomorphisms that will 
be our primary case of interest in this review. 
We define a Yang-Mills - scalar fields action functional as 
\begin{equation} \label{action}
S= \frac{1}{g^{2}}{\rm Tr} \Bigl( \frac{1}{4} F_{jk}F^{jk} + \frac{1}{2}\sum_{\alpha}[\nabla_{j}, 
\Phi^{\alpha}][\nabla^{j}, \Phi^{\alpha}] + 
V(\Phi ) \Bigr)
\end{equation}
where $\Phi^{\alpha}$, $\alpha = 1, \dots, n$  are endomorphisms representing $n$ scalar fields, 
$V(\Phi)$ is a potential that we will assume to be a polynomial. It is easy to introduce fermionic  fields, but 
for now we will concentrate on bosonic systems. 

The equations  of motion corresponding to  functional (\ref{action}) read
\begin{eqnarray} \label{eqofmo}
&& [\nabla^{j}, [\nabla_{j}, \nabla_{k}]] = - \sum_{\alpha} \Phi^{\alpha} [\nabla_{k}, \Phi_{\alpha}] \, , \nonumber \\
&&  [\nabla^{j}, [\nabla_{j}, \Phi_{\alpha}]] = \partial_{\alpha} V (\Phi )
\end{eqnarray}  
where $\partial_{\alpha} V$ stands for the partial derivative of the potential with respect to the $\alpha$-th variable.  

Consider now  scalar fields $\Psi_{\alpha}$ ``in the fundamental representation'', that is 
$\Psi_{\alpha}\in E$ are elements of the module $E$ itself. In that case the action functional analogous to 
(\ref{action}) can be constructed  with a help of the ${\cal A}$-valued inner product on $E$ introduced in section \ref{innpr}.
Namely the action functional reads 
\begin{equation}
S = \frac{1}{g^{2}}{\rm Tr} \Bigl( \frac{1}{4} F_{jk}F^{jk} + 
\frac{1}{2}\sum_{\alpha}<\nabla_{j} \Psi^{\alpha}, \nabla^{j} \Psi^{\alpha}>_{\cal A} + 
V(<\Psi_{\alpha}, \Psi_{\beta}>_{\cal A} ) \Bigr) \, . 
\end{equation}

\section{Solitons}
\subsection{Finite energy solutions} \label{exact} 
We can look for solutions to the equations of motion (\ref{eqofmo}) that have a finite action, i.e. the 
functional  (\ref{action}) is well defined and finite. We will  refer to this type of solutions as 
solitons.  On a general module  $E= {\cal F}^{k}\oplus (\tilde {\cal A})^{N}$ the scalar fields $\Phi_{\alpha}$ 
and the Yang-Mills field $\nabla_{i}$ can be decomposed according to (\ref{endomorphism}) and (\ref{connection}) 
respectively. Let us first consider the case of a single scalar field $\Phi$. Let $\{\phi_{0}, \dots , \phi_{p}\}$ be 
the set of extrema of the function $V(\Phi)$. Also assume that  $V(\phi_{0})=0$. Then we have a set of simple 
finite action solutions to (\ref{eqofmo}) constructed as follows. The solutions have a block-diagonal form 
with respect to the decomposition  $E= {\cal F}^{k}\oplus (\tilde {\cal A})^{N}$. The Yang-Mills fields are 
\begin{equation} \label{con_s}
\nabla_{j} = \left( 
\begin{array}{cc}  i\theta_{jk}^{-1}\hat x^{k} & 0\\
0&\partial_{j} \end{array} \right) + i \left( 
\begin{array}{cc}
(D_{j})_{i}^{l}   &  0  \\
  0  & (\hat D_{j})^{\alpha}_{\beta} \end{array} \right) 
\end{equation}
where 
\begin{equation}\label{D}
(D_{j})_{i}^{l} = 
\left( \begin{array}{ccc} 
d_{j(1)}& \dots & 0 \\
0 & \ddots & 0 \\
0 & \dots & d_{j(k)}  
\end{array} \right) \, , \qquad 
(\hat D_{j})_{\alpha}^{\beta} = 
\left( \begin{array}{ccc} 
a_{j(1)}\cdot {\bf 1}& \dots & 0 \\
0 & \ddots & 0 \\
0 & \dots & a_{j(N)} \cdot {\bf 1}  
\end{array} \right)
\end{equation}
and $d_{j(i)}$, $a_{j(\alpha)}$ are all numbers.
The scalar field reads 
\begin{equation} \label{scalar}
\Phi = \left( \begin{array}{cc} 
\left( \begin{array}{ccc} 
\lambda_{1}& \dots & 0 \\
0 & \ddots & 0 \\
0 & \dots & \lambda_{k}  
\end{array} \right) & 0 \\
0 & \phi_{0} \cdot {\bf 1} 
\end{array} \right) 
\end{equation}
where each $\lambda_{i}$ is an element of  the set of extrema $\{\phi_{i} \}$.
Substituting these solutions into (\ref{action}) and using the definition of trace (\ref{Trace3}) 
one finds the value of the soliton action 
\begin{equation} 
S = \frac{|{\rm Pfaff}( 2\pi \theta)| }{g^{2}}(\sum_{1}^{k} V(\lambda_{i}) + \frac{k}{4}  
(\theta^{-1})_{in}g^{ij}g^{nl}(\theta^{-1})_{jl}) \, .  
\end{equation}
One sees from this expression that the moduli  $d_{i}$, $a_{\alpha}$  are zero modes of  the solitons. 


\subsection{Partial isometries} \label{PI}
In the previous section we constructed finite energy solutions to the equations of motion by starting with 
a module ${\cal F}^{k}\oplus (\tilde {\cal A})^{N} $. We think that this approach is more in 
the spirit of noncommutative geometry than the  ideology accepted in the  literature  based on partial isometry operators and 
most clearly stated in  \cite{Exact}. 
In this section we would like to explain the connection between the two pictures. 

This connection is based on the remark that ${\cal F}^{k}\oplus ( {\cal A})^{N} $ considered as an $\cal A$ 
module is isomorphic to $ ( {\cal A})^{N} $. This isomorphism permits us to represent a very simple solution 
on  ${\cal F}^{k}\oplus ( {\cal A})^{N} $ as a more complicated solution on a simpler module  $ ( {\cal A})^{N} $. 
A rigorous proof of this isomorphism will be given in section \ref{math}. Here we will present a very transparent 
but not completely rigorous consideration 
 restricting ourselves for simplicity to the case $N=1$.

Notice that a free module $({\cal A})^{1}$ can be decomposed into a direct sum of countably many copies of module ${\cal F}$. 
This can be done in the following way. As it was explained in section \ref{algebras} each algebra $\cal A$ acts 
on a subset of ${\cal H}=L_{2}({\mathbb R}^{d/2})$ consisting of Schwartz class functions. Let $e_{i}$, $i\in {\mathbb N}$ be 
a countable basis in $\cal H$ such that each $e_{i}$ belongs to the Schwartz space. For example one can take the basis 
consisting of energy eigenfunctions of a $d$-dimensional harmonic oscillator. 
Then the rank one free module can be decomposed as 
$$
({\cal A})^{1} = \bigoplus_{i=0}^{\infty}  P_{i} \cdot {\cal A}
$$
where $P_{i} = |e_{i}\rangle \langle e_{i}|$ are orthogonal projectors onto the one-dimensional subspaces spanned by 
the vectors $e_{i}$. Each submodule $ P_{i} \cdot {\cal A}$ is isomorphic to $\cal F$. The isomorphism is established 
by means of the mapping 
$$
P_{i}\cdot f \mapsto \langle e_{i} | f \in {\cal S}({\mathbb R}^{d/2}) \cong {\cal F} \, .
$$

As a consequence of the above decomposition 
 the modules  ${\cal F}^{k}\oplus ( {\cal A})^{1} $ and   $({\cal A})^{1}$ are isomorphic.
Using this isomorphism we  construct a solution to the Yang-Mills-Higgs equations of motion on  $( {\cal A})^{1} $ 
by first taking the solutions on  ${\cal F}^{k}\oplus ( {\cal A})^{1} $ described in the previous section and 
then applying the isomorphism.  
To this end let us first choose a particular isomorphism. This can be done by choosing a basis 
$e_{i}\in {\cal S}({\mathbb R}^{d/2}) \subset {\cal H}$ in such a way that ${\cal F}^{k}$ sitting inside  $( {\cal A})^{1} $ 
is identified with $\oplus_{i=0}^{k-1}P_{i}\cdot {\cal A}\equiv \Pi_{k} \cdot {\cal A}$, $\Pi_{k}=\oplus_{i=0}^{k-1}P_{i}$. 
Define an operator 
$$
S = \sum_{i=0}^{\infty} |e_{i+1}\rangle \langle e_{i}| :  {\cal H} \to {\cal H}  \, . 
$$ 
It can be easily checked to satisfy 
\begin{equation} \label{SS}
(S^{\dagger})^{k}S^{k} = 1 \, , \qquad  S^{k}(S^{\dagger})^{k} = 1 - \Pi_{k} \, . 
\end{equation}
An operator satisfying such a property is called a partial isometry. 

The desired isomorphism is established by a mapping $M:{\cal F}^{k}\oplus ( {\cal A})^{1}\to  ({\cal A})^{1}$ defined as
$$
M : ( \sum_{i=0}^{k-1}\langle e_{i} |f , g) \mapsto \Pi_{k}\cdot f + S^{k}\cdot g \, . 
$$
The inverse mapping is  
$$
M^{-1}: f \mapsto ( \sum_{i=0}^{k-1}\langle e_{i} |f , (S^{\dagger})^{k}\cdot f) \, . 
$$
Here the dot ``$\cdot$'' stands for the composition of operators. 

Let us first find the image 
$\tilde \nabla_{j}^{0}: ( {\cal A})^{1} \to ( {\cal A})^{1}$
of the standard connection (\ref{st_conn}) under this isomorphism. We have  
$$
\tilde \nabla_{j}^{0}(f) = M \nabla^{0}_{j}M^{-1}(f) = i\Pi_{k}\cdot f \cdot \theta^{-1}_{jl}\hat x^{l} + 
S^{k}\partial_{j}(S^{\dagger})^{k} f \, . 
$$
Using the identity 
\begin{equation} \label{id}
-i[\theta^{-1}_{jl}\hat x^{l}, f] = \partial_{j}(f)
\end{equation}
we obtain from the above expression
\begin{equation} \label{con2}
\tilde \nabla_{j}^{0} = \partial_{j} + S^{k}[\partial_{j}, (S^{\dagger})^{k}] + i \theta^{-1}_{jl}\Pi_{k}\cdot \hat x^{l} \, . 
\end{equation}
It is also easy to check that under the isomorphism $M$ the gauge field (\ref{D}) and 
the scalar field (\ref{scalar}) give rise to the following operators on ${\cal A}^{1}$
\begin{eqnarray} \label{A2}
&& \tilde A_{j} = \sum_{i=0}^{k-1}d_{j(i+1)}|e_{i}\rangle \langle e_{i}| + a_{j}(1 -\Pi_{k}) \, , \\
&& \tilde \Phi =  \sum_{i=0}^{k-1}\lambda_{i+1}|e_{i}\rangle \langle e_{i}| + \phi_{0}(1-\Pi_{k})\, .  
\end{eqnarray}
Thus the whole connection (\ref{con_s}) on ${\cal F}^{k}\oplus (\tilde {\cal A})^{N}$
 corresponds to a connection $\tilde \nabla_{j}$ on ${\cal A}^{1}$ 
of the form $\tilde \nabla_{j} = \tilde \nabla_{j}^{0} +  \tilde A_{j}$ where  $\tilde \nabla_{j}^{0}$, $\tilde A_{j}$ 
are given by expressions (\ref{con2}), (\ref{A2}).

To compare these expressions with the ones obtained in    \cite{Exact} whose  conventions are most widely used in 
the literature we first rewrite the connection (\ref{con2}) 
in the complex coordinates (\ref{cc}): 
$$
\tilde \nabla_{z_{\alpha}}^{0} = \partial_{z_{\alpha}} + S^{k}[ \partial_{z_{\alpha}},  (S^{\dagger})^{k}] + 
\theta_{\alpha}^{-1/2}\Pi_{k}\cdot a_{\alpha}^{\dagger}
$$ 
where for simplicity we assumed all $\theta_{\alpha}>0$. Using identities (\ref{SS}) and (\ref{id})) we can rewrite it as 
$$
\tilde \nabla_{z_{\alpha}}^{0} = \partial_{z_{\alpha}}  - 
\theta_{\alpha}^{-1/2}( S^{k}\cdot a_{\alpha}^{\dagger} \cdot (S^{\dagger})^{k} - a_{\alpha}^{\dagger}) \, . 
$$
Now the case considered in \cite{Exact} corresponds to $d=2$, $\lambda_{i} = \phi_{\star}$, $\phi_{0}=0$. Then the 
solution with $d_{j(i+1)}=0$, $a_{j}=0$  that in the language of \cite{Exact} 
corresponds to dressing a trivial gauge field by a partial isometry $S^{k}$
has the form 
\begin{eqnarray*}
&& \tilde \nabla_{z}^{0} = \partial_{z} + \theta^{-1/2}( a^{\dagger} - C) \, , \\
&& \tilde \nabla_{\bar z}^{0} = \partial_{\bar z} + \theta^{-1/2}( a - C^{\dagger}) \, , \\
&& \tilde \Phi = \phi_{\star}(1 - \Pi_{k}) 
\end{eqnarray*}
where 
\begin{equation} \label{C}
C = S^{k}  a^{\dagger}  (S^{\dagger})^{k} \, , \qquad C^{\dagger} =  S^{k}  a  (S^{\dagger})^{k}\, . 
\end{equation}
These expressions  up to some minor differences in notations coincide with the ones obtained  in \cite{Exact}.

We feel that the following pedagogical remark is in order here. Although the ideology of doing ``dressing transformations'' 
by partial isometries (\ref{SS}) to generate new solutions of Yang-Mills-Higgs equations (\ref{eqofmo}) may 
seem quite appealing in its simplicity, it should be taken with some care. Thus if one regards  (\ref{eqofmo}) merely 
as an algebraic equation satisfied by operators $\nabla_{j}$, $\Phi^{\alpha}$ then naively one may wish to 
generate new solutions by simply taking a transformation 
$$
\nabla_{j} \mapsto \nabla_{j}' =  S^{k}  \nabla_{j}  (S^{\dagger})^{k} \, , \qquad 
\Phi^{\alpha} \mapsto {\Phi^{\alpha}}' =  S^{k}  \Phi^{\alpha}  (S^{\dagger})^{k} \, . 
$$
Although formally this transformation results in operators $ \nabla_{j}'$, ${\Phi^{\alpha}}'$ satisfying  (\ref{eqofmo}) 
it cannot be considered as a new solution to  the Yang-Mills-Higgs equations  of motion because  the operators $ \nabla_{j}'$
fail to satisfy Leibniz rule and thus are not Yang-Mills fields any more. The construction of \cite{Exact} circumvents 
this problem essentially by a trick, introducing fields $C$, $C^{\dagger}$ that are endomorphisms and performing a dressing 
transformation (\ref{C}) on these fields and not on $\nabla_{j}$'s directly. 


\subsection{Scalar field solitons in the $\theta \to \infty$ limit} 
Consider  a single scalar field $\Phi$ on a module ${\cal A}^{N}$ with action functional 
\begin{equation}
S = \frac{1}{g^{2}}\int d^{d}x ( \frac{1}{2}
[\partial_{i}, \Phi][\partial^{i}, \Phi] + V(\Phi ) ) 
\end{equation}
where the function $V(x)$ is assumed to be a  polynomial,
 $\Phi \in End_{\cal A} {\cal A}^{N}$ is  Hermitian and $g$ is a coupling constant. 
For simplicity we will assume ${\cal A} = {\cal S}({\mathbb R}_{\theta}^{d})$ and 
that $\theta^{ij}$ is brought to the standard form (\ref{standard_theta}). Moreover we will take 
$\theta_{1} = \theta_{2}=\dots =\theta_{d/2}=\theta >0$ so that noncommutativity is parameterized by a single parameter 
$\theta$. 
  Following \cite{NCsol} let us further rescale   
 the coordinates $x^{i}$ on  ${\mathbb R}_{\theta}^{d}$ as $x^{i} \mapsto x^{i}\sqrt{\theta }$. Rescaling also 
the coupling constant as $g^{2} \mapsto g^{2}/ \theta^{d/2}$ we obtain the action functional 
\begin{equation} \label{S'}
S' = \frac{1}{g^{2}}\int d^{d}x ( 
\frac{1}{2\theta} [\partial_{i}, \Phi][\partial^{i}, \Phi] + V(\Phi ) ) \, . 
\end{equation}
Now the operators corresponding to rescaled coordinates all satisfy the canonical commutation relations 
\begin{equation} \label{ccr'}
[\hat x^{j}, \hat x^{j+n}] = i \, , \enspace  j=1, \dots , n 
\end{equation}
and $\theta$ enters (\ref{S'}) as a formal parameter. We see that taking the limit $\theta \to \infty$ results in 
dropping the kinetic term from (\ref{S'}) while the algebra we work with stays the same specified by (\ref{ccr'}). 
Thus when $\theta \to \infty$ the approximate equations of motion take a very simple form  
\begin{equation} \label{eq}
 V'(\Phi ) = 0  
\end{equation}
where $V'$ stands for the derivative of  $V(x)$.
For a polynomial potential $V(x)$  it is easy to find the complete set of solutions to (\ref{eq}). 
 Let $\phi_{1}, \dots , \phi_{p}$ be a set of zeroes of $V'(x)$. Considering $\Phi$ as a pseudodifferential operator 
acting in Hilbert space $\cal H$ we immediately imply from (\ref{eq}) that any eigenvalue of operator $\Phi$ should 
coincide with one of the numbers $\phi_{i}$. Therefore we can represent $\Phi$ as 
\begin{equation} \label{Phi}
\Phi = \sum_{i=1}^{p} \phi_{i}P_{i}
\end{equation}
where $P_{i}$ are orthogonal projectors on  subspaces  ${\cal H}_{i}$ such that ${\cal H}=\oplus_{i=1}^{p}{\cal H}_{i}$. 
The energy of solution (\ref{Phi}) in the limit $\theta \to \infty$ when we can neglect the kinetic term is given by 
$$
E = \frac{1}{g^{2}} \sum_{i=1}^{p} {\rm Tr P_{i}} V(\phi_{i}) = \frac{1}{g^{2}} \sum_{i=1}^{p} {\rm dim }{\cal H}_{i}
V(\phi_{i}) \, . 
$$ 
From this expression we see that the energy is finite provided  the dimensions $  {\rm dim }{\cal H}_{i}$ 
are infinite only when the corresponding $V(\phi_{i})$ vanish. This in its turn may be possible only when 
the set of extrema of $V(x)$ and the set of its zeroes has a nontrivial intersection. It is convenient 
to choose $x=0$ to be such a point. That means that $V(x)$ is  of the form 
$V(x)=x^{2}\tilde V(x)$ where $\tilde V(x)$ is an arbitrary polynomial. In the case when $x=0$ is the only such point 
an arbitrary (asymptotic) finite energy solution can be written as 
$\Phi = \sum_{i=1}^{p}\phi_{i}P_{i}$ where ${\rm Tr}P_{i}<\infty$ for all $i$.

It is not hard to implement first order corrections induced by nonzero kinetic term. 
We can write an expansion for the exact solution 
$$
\Phi = \Phi_{0} + \frac{1}{\theta}\Phi_{1} + \dots 
$$
where $\Phi_{0}$ has the form  (\ref{Phi}).
The first order correction to the energy is then given by 
$$
K\equiv \frac{1}{2\theta g^{2}}{\rm Tr} [\partial_{i}, \Phi_{0}]  [\partial^{i}, \Phi_{0}] \, . 
$$

Thus we   may seek an improved solution by minimizing the kinetic 
energy over all zero order solutions  $\Phi_{0}$  \cite{NCsol}, \cite{Ganor_etal}, \cite{Gop}. 
For simplicity let us consider the case when 
$\Phi_{0}$ is of the form 
$$
\Phi_{0} = \phi_{1}P = \phi_{1}\sum_{i=0}^{k}|e_{i}\rangle\langle e_{i}|
$$
where $|e_{i}\rangle \in {\cal H}$ is a finite orthonormal system of vectors. Substituting it in the kinetic energy 
and employing complex coordinates (\ref{cc}) we 
obtain 
\begin{equation} \label{K}
 K= -\frac{1}{2\theta g^{2}}\sum_{\alpha=1}^{d/2} {\rm Tr} [a_{\alpha}, \Phi_{0}] [a_{\alpha}^{\dagger}, \Phi_{0}]= 
\frac{\phi_{1}^{2}}{\theta g^{2}}\sum_{\alpha=1}^{d/2}{\rm Tr} \Bigl( F^{\dagger}_{\alpha}F_{\alpha} + P \Bigr)
\end{equation}
where 
$$
F_{\alpha} = (1-P)a_{\alpha}P \, .
$$
And we see from this expression that the kinetic energy satisfies a Bogomolnyi type bound 
$$
K\ge \frac{\phi_{1}^{2}kd}{2\theta g^{2}} 
$$
that is saturated by the projectors $P$ satisfying $(1-P)a_{\alpha}P=0$ that means that $P$ projects on a subspace 
invariant under annihilation operators $a_{\alpha}$. Such subspaces are spanned by coherent states 
$$
|z \rangle = e^{z^{\alpha} a^{\dagger}_{\alpha}}|0\rangle 
$$
that are eigenstates of operators  $a_{\alpha}$ with eigenvalues $z^{\alpha}\in {\mathbb C}$. The numbers $z^{\alpha}$
can be considered as complex coordinates on the moduli space of approximate (first order in $1/\theta$) 
solutions to the equation of motion. 
Summarizing the above discussion we can say that strictly at the limit $\theta = \infty$ we have general solutions 
to the equation of motion of the form (\ref{Phi}) labelled by a collection of $p$ orthogonal 
projectors $P_{i}$ on finite-dimensional 
subspaces. Each such projector of rank $k$ specifies a point in an infinite-dimensional 
    Grassmannian $Gr(k, {\cal H})$, i.e., the corresponding moduli space of solutions is infinite-dimensional 
(it is natural to identify solutions related by a unitary transformation). 
However if one takes into account finite $\theta$ corrections the corresponding moduli space reduces to a finite-dimensional 
complex manifold.  

It was argued in the original paper \cite{NCsol} that the exact solitonic solutions in scalar theories exist 
at finite $\theta$. Recently the existence of rotationally invariant 
solitons at finite  $\theta$ was proved rigorously in \cite{existence_stab}. 
Namely the authors of  \cite{existence_stab} obtained the following result. Let $P_{i}$ denote the  
orthogonal projectors onto the eigenspaces of the number operator 
$$
N=\sum_{\alpha=1}^{d/2}a^{\dagger}_{\alpha}a_{\alpha}
$$ of the $d/2$-dimensional harmonic 
oscillator. Then {\it for any projection $P$ on ${\cal H}\cong L_{2}({\mathbb R}^{d/2})$, which is the sum 
of a finite number of projections $P_{i}$, there is a unique family $\Phi_{\theta}$ of rotationally invariant 
solutions to the equation of motion 
$$
[\partial_{i}, [\partial^{i}, \Phi]] = \theta V'(\Phi )
$$
that depend smoothly on $\theta$ and satisfy $V''(\Phi_{\theta})>0$ and $\Phi_{\theta} \to \phi \cdot P$ in Hilbert-Schmidt operator 
norm as $\theta \to \infty$.} Here $\phi$ stands for the local minimum of $V(x)$. The theorem is proved in the assumption that $V$ 
has exactly one local minimum and one local maximum (see \cite{existence_stab} for details).


A similar construction to (\ref{Phi}) also applies to the case of noncommutative tori that were discussed at length in the first 
part of this review. In that case we may consider a single scalar field $\Phi$ taking values in 
endomorphisms  of a free module $( T_{\theta}^{d})^{N}$, i.e., $\Phi$ is an  $N\times N$ 
matrix with entries from $T_{\theta}^{d}$. The action functional is 
\begin{equation} \label{act_tor}
\frac{1}{g^{2}}{\rm Tr}( [\partial_{i}, \Phi][\partial^{i}, \Phi] + V(\Phi ) ) 
\end{equation}
where $\partial_{i}$ is the standard derivative on $E$ and ${\rm Tr}$ is the canonical trace on  the algebra 
of endomorphisms 
(see part I for precise definitions and discussion). The matrix  $\theta^{ij}$ specifying the noncommutative torus 
$T_{\theta}^{d}$ is defined up to addition of an antisymmetric matrix with integer entries.  Therefore the limit 
$\theta \to \infty$ strictly speaking does not make sense. We may still however neglect the kinetic term in (\ref{act_tor}) 
in the limit of large volume  (terms entering the kinetic part scale as $1/R_{i}^{2}$ where $R_{i}$ are radii of the torus).
In that limit we obtain approximate solutions to the equation of motion of the form (\ref{Phi}) where 
$P_{i}$ now are mutually orthogonal projectors, i.e., elements of $Mat_{N}(T_{\theta}^{d})$ satisfying 
$P_{i}P_{j}=\delta_{ij}P_{i}$, $P^{\dagger}_{i}=P_{i}$. Any solution of the form $\Phi = \sum_{i=1}^{p} \phi_{i}P_{i}$ 
will have a finite energy 
and there is no need for additional assumptions on the form of potential $V(x)$ that we had to impose in the 
${\mathbb R}^{d}_{\theta}$ case. 
In the two-dimensional case one can employ the Powers-Rieffel  general construction of projector operators (for example 
see \cite{Connes_book}, \cite{GBondia_etal}). However the value of the soliton energy can be calculated without knowing 
the detailed form of the projectors: 
$$
E = \frac{1}{g^{2}}\sum_{i=1}^{p}V(\phi_{i}){\rm Tr}P_{i} = \frac{1}{g^{2}}\sum_{i=1}^{p}V(\phi_{i}) {\rm ch_{0}}(P_{i})
$$     
where ${\rm ch_{0}}(P_{i})$ stand for the zero order Chern numbers (dimensions) of the projective modules specified by 
$P_{i}$. As we explained in the first part of the review the Chern numbers $ch_{n}({\cal E})$ of an arbitrary 
 projective module $\cal E$ 
depend only on the K-theory class of $\cal E$ and the noncommutativity matrix $\theta^{ij}$. The K-theory class 
in its turn is specified by topological integers. The explicit formulas  for ${\rm dim {\cal E}} = {\rm Tr} P$ via 
these numbers are given for $d=2,3,4$-dimensional tori in section 6.7. For example in the case $d=4$ we may write 
the energy of the soliton corresponding to a solution $\Phi = \phi_{1}P$ as  
$$
E= \frac{V(\phi_{1})}{g^{2}}\Bigl( n + \frac{1}{2}{\rm tr}(\theta m) + q{\rm Pfaff}(\theta ) \Bigr) \, . 
$$
Here $n, m_{ij}, q$ are topological integers characterizing the K-theory class of $E=P{\cal A}^{1}$ corresponding 
in string theory language to the numbers of $D4$, $D2$ and $D0$-branes wrapped on $T^{4}_{\theta}$.

To conclude this section let us remark that the approximate solitons  of this section and the exact ones discussed in 
sections \ref{exact}, \ref{PI} found an important application in describing D-branes as solitonic solutions of the tachyon field 
in the presence of large $B$-field \cite{Dasgupta_etal}, \cite{Harvey_tach} (see also the annotated list of
 references in the last section of this paper).

\section{Instantons}
\subsection{Instanton equation on ${\cal F}^{k}\oplus \tilde \Gamma_{\theta}^{N}$ modules} \label{inst1}
In the part one of the review we discussed instantons on noncommutative tori and their interpretation as 
 BPS fields; see Sec. 8.1, 8.2.  Here we would like to consider instantons on noncommutative ${\mathbb R}^{4}$ space.

In the above discussion of solitons  
we used the algebra ${\cal A} = \tilde {\cal S}({\mathbb R}^{4}_{\theta})$. This is appropriate because the corresponding 
solutions rapidly decrease at infinity. Instantons do not have this property. Therefore we should replace 
$\tilde {\cal S}({\mathbb R}^{d}_{\theta})$ with something else. It turns out that the appropriate algebra is 
$\cup_{m< -1}\Gamma^{m}({\mathbb R}^{4}_{\theta} )\equiv \Gamma_{\theta}$. Notice that this algebra could be also used in the 
analysis of solitons.

The Fock module $\cal F$ can be considered also as a projective $\tilde \Gamma_{\theta}$-module. Thus every projective 
$\tilde \Gamma_{\theta}$-module is isomorphic to one of the modules ${\cal F}_{k, N} = {\cal F}^{k}\oplus \tilde \Gamma_{\theta}^{N}$.
Introducing  complex coordinates
\begin{equation}\label{complex_str}
z_{1} = \frac{1}{\sqrt{2}}(x^{1} + ix^{3}) \, , \qquad z_{2} = \frac{1}{\sqrt{2}}(x^{2} + ix^{4})
\end{equation}
we have 
\begin{eqnarray*}
&& D_{1} = \frac{1}{\sqrt{2}}(\nabla_{1} - i \nabla_{3}) \, , \qquad   
D_{2} = \frac{1}{\sqrt{2}}(\nabla_{2} - i \nabla_{4})\, , \\
&& D_{1}^{\dagger} = \frac{1}{\sqrt{2}}(\nabla_{1} +i \nabla_{3}) \, , \qquad   
D_{2}^{\dagger} = \frac{1}{\sqrt{2}}(\nabla_{2} +  i \nabla_{4})
\end{eqnarray*}
where $D_{i}, D^{\dagger}_{i}$ stand for  $\nabla_{z_{i}}$,  $\nabla_{\bar z_{i}}$.
In these coordinates  the antiinstanton (self-duality) equation\footnote{Note that our choice of complex structure 
(\ref{complex_str}) is different from the one employed in \cite{Donaldson}, \cite{NS}, \cite{revisited}, 
\cite{Trieste}, and although the equation looks formally the same it corresponds to an antiinstanton versus 
an instanton considered in those papers.}
$F_{ij} = \frac{1}{2}\epsilon_{ijkl}F^{kl}$ takes the form 
\begin{equation} \label{inst_hol}
 [D_{1}, D_{2}] = 0 \, , 
\end{equation}
\begin{equation} \label{inst_real}
 [D_{1}, D_{1}^{\dagger}] + [D_{2}, D_{2}^{\dagger}] = 0 \, . 
\end{equation}
We also assume that in our set of coordinates $x^{j}$ the matrix $\theta^{ij}$ has the standard form (\ref{standard_theta})
\begin{equation}
\theta^{ij} = \left( \begin{array}{cccc}
0&0&\theta_{1}&0\\
0&0&0&\theta_{2}\\
-\theta_{1}&0&0&0\\
0&-\theta_{2}&0&0
\end{array} \right) \, . 
 \end{equation}
There are two distinct cases 
${\rm Pfaff}(\theta)= \theta_{1}\cdot \theta_{2} >0$ and  $\theta_{1}\cdot \theta_{2} <0$.  
These two cases are connected by an orientation-reversing reflection.  That reflection also interchanges instanton and 
antiinstanton solutions. Without loss of generality we will limit ourselves to the case of the antiinstanton equation and 
will allow  both signs of the Pfaffian.

As explained in section \ref{conn} the connections $D_{1}$, $D_{2}$ have a general form 
$$
D_{1} = D_{1}^{0} + \left( \begin{array}{cc} 
(B_{1})_{r}^{s} & \langle I_{r}^{\beta} | \\
|K_{\alpha}^{s}\rangle & (\hat R_{1})_{\alpha}^{\beta}   
\end{array} \right) \, , \quad 
D_{2} = D_{2}^{0} + \left( \begin{array}{cc} 
(B_{2})_{r}^{s} & \langle L_{r}^{\beta} | \\
|J_{\alpha}^{s}\rangle & (\hat R_{2})_{\alpha}^{\beta}   
\end{array} \right) 
$$ 
where the indices $r,s$ run from 1 to $k$ and $\alpha, \beta$ run from 1 to $N$. Below for brevity we will 
suppress these indices.  Standard fiducial connection $D_{i}^{0}$ is given by (\ref{complex_conn}). 
To shorten the notations we write 
$$
D_{i}^{0} = \left( \begin{array}{cc} 
\tilde \partial_{i} & 0 \\
0& \partial_{i} \end{array} \right) \, . 
$$
The curvature tensor $-i\theta^{-1}_{jk}$ of $\tilde \partial_{i}$ can be decomposed into selfdual and antiselfdual parts 
according to 
$$
\theta^{-1}_{s.d.} = \frac{1}{2}(\theta^{-1} + \tilde \theta^{-1}) = 
  \left( \begin{array}{cc}
0 & f^{+}\\
-f^{+}& 0
\end{array} \right) \, , \quad  \theta^{-1}_{a.s.d.} = \frac{1}{2}(\theta^{-1} -  \tilde \theta^{-1})
 \left( \begin{array}{cc}
0 & f^{-}\\
-f^{-} & 0
\end{array} \right) \, . 
$$
where
$$
f^{+} = \left( \begin{array}{cc}
\theta_{2}^{-1} - \theta_{1}^{-1} &0 \\
0 & \theta_{1}^{-1} - \theta_{2}^{-1}
\end{array} \right) \, , \quad 
f^{-} = \left( \begin{array}{cc}
-(\theta_{1}^{-1} + \theta_{2}^{-1}) &0 \\
0& -(\theta_{1}^{-1} + \theta_{2}^{-1})
\end{array} \right) 
 $$
and $\tilde \theta_{ij} =\frac{1}{2}\epsilon_{ijkl}\theta_{kl}^{-1}$.

Assuming $K=L=0$, that can be considered as  gauge conditions, the first  instanton equation (\ref{inst_hol}) 
gives rise to the following set of equations 
\begin{eqnarray}\label{matr1} 
 && [B_{1}, B_{2} ] + \langle I|J\rangle = 0 \, , \\
&& \langle I|\hat R_{2} - B_{2}\langle I| + \langle I|\partial_{2} - \tilde \partial_{2}\langle I| = 0 \, , \nonumber \\
&& \hat R_{1}|J\rangle -|J\rangle B_{1} + \partial_{1}|J\rangle - |J\rangle\tilde \partial_{1} = 0 \, , \nonumber \\
&& [\hat R_{1}, \hat R_{2}] - |J\rangle\langle I| + [\partial_{1}, \hat R_{2}] + [\hat R_{2}, \partial_{2}] = 0\, \nonumber .  
\end{eqnarray}
The second (real) instanton equation (\ref{inst_real}) can be also represented in terms of four equations for 
each block. Here we write down only equation for the upper left block:
\begin{equation} \label{matr2}
-(\theta_{1}^{-1} + \theta_{2}^{-1}) + [B_{1}, B_{1}^{\dagger}] +  [B_{2}, B_{2}^{\dagger}] + 
\langle I | I\rangle - \langle J | J \rangle = 0 \, . 
\end{equation}
It is easy to see that equations (\ref{matr1}), (\ref{matr2}) closely resemble the ADHM matrix equations.
Moreover let us take
 an ansatz 
$$
\langle I_{r}^{\beta}| = (I^{0})_{r}^{\beta} \langle \Phi|\, , \qquad  
|J_{\alpha}^{s}\rangle = (J^{0})^{s}_{\alpha} |\Psi\rangle
$$
 where $ (I^{0})_{r}^{\beta}$, $(J^{0})^{s}_{\alpha}$ 
are $N\times k$ and $k\times N$ matrices with numerical entries and $\langle \Phi|$, $ |\Psi\rangle$ are 
such that $\langle \Phi |\Psi\rangle=1$. Then it follows from (\ref{matr1}), (\ref{matr2}) that 
the  matrices $B_{i}$, $I^{0}$, $J^{0}$  satisfy the noncommutative ADHM 
equations 
\begin{equation} \label{ADHM1}
 [B_{1}, B_{2}] + I^{0}J^{0} = 0 \, , 
\end{equation}
\begin{equation}\label{ADHM2}
 [B_{1}, B_{1}^{\dagger}] +  [B_{2}, B_{2}^{\dagger}] +  I^{0} (I^{0})^{\dagger} - (J^{0})^{\dagger}J^{0} = 
\zeta_{r}
\end{equation}
where $\zeta_{r}=\theta_{1}^{-1} + \theta_{2}^{-1}$.

It looks like a natural conjecture that once a solution 
to the ADHM equations  (\ref{ADHM1}), (\ref{ADHM2}) is chosen one can find $\langle \Phi|$, $ |\Psi\rangle$, $\hat R_{1}$ 
$\hat R_{2}$ such that the rest of the equations following from (\ref{inst_hol}), (\ref{inst_real}) is satisfied.

Thus we see that following our approach based on  classification of projective modules ${\cal F}_{k, N}$ 
we obtained the noncommutative  ADHM equations in a seemingly effortless way. The relation between this approach 
and the standard one of \cite{NS}, \cite{revisited}  can be established in a similar way to how  it was done for solitons 
in section \ref{PI}. See sections \ref{ADHM22}, \ref{inst2} for a more detailed explanation of this relation and a more 
detailed treatment of the ADHM construction. 


\subsection{Hypoelliptic operators, ${\cal F}_{k,N}$-modules and partial isometries} \label{math}
In this section we will formulate some mathematical results that are needed for a more complete and more rigorous
treatment of gauge theory on noncommutative ${\mathbb R}^{d}$.

In section \ref{algebras} we introduced a class of functions 
$  \Gamma^{m}_{\rho}\equiv \Gamma^{m}_{\rho}({\mathbb R}^{d})$. Let us introduce 
a subclass $\Sigma^{m}_{\rho}\subset \Gamma^{m}_{\rho}$ consisting of smooth functions 
$f(x)$ that  satisfy 
\begin{equation} \label{sigma_space}
|f(x)|\le C \ | x \|^{m} \, , \qquad  
|\partial_{\alpha}f(x)|\le C_{\alpha}|f(x)|  \| x \|^{ m - \rho|\alpha|}  
\end{equation}
for  all $\|x\|>R$ for some $R>0$, i.e. outside of a sphere of sufficiently large radius.
We say that a function $f(x)$ belongs to the class ${\rm H}\Gamma^{m, m_{0}}_{\rho}$ if $f\in \Sigma^{m}_{\rho}$ 
and there exists a function $g(x)\in  \Sigma^{-m_{0}}_{\rho}$ such that $f(x)g(x)=1$ for $\|x\| > r$ where 
$r$ is some positive number. In other words $f(x)$  must have an  inverse 
 outside of a sphere of sufficiently large radius 
with respect to the usual pointwise multiplication  and this inverse must be a function from 
 $ \Sigma^{-m_{0}}_{\rho}$. One can prove (see for example \cite{Shubin}) that given a function 
$f\in {\rm H}\Gamma^{m, m_{0}}_{\rho}$ there exists a function $g_{\theta}\in {\rm H}\Gamma^{-m, -m_{0}}_{\rho}$
called parametrix that satisfies 
$$
1 - f\ast g_{\theta} \in  {\cal S}({\mathbb R}^{d}) \, , \qquad 1 - g_{\theta}\ast f \in  {\cal S}({\mathbb R}^{d}) 
$$
where $\ast$ stands for the star product defined by $\theta$. Essentially what we are saying here is that 
 if a function $f(x)$ is invertible 
up to an element from $ {\cal S}({\mathbb R}^{d})$ (invertible at large distances) with respect to the pointwise 
product it is also invertible up to an element from  $ {\cal S}({\mathbb R}^{d})$ with respect to the star product.
Note that if there is another element $g_{\theta}'\in {\rm H}\Gamma^{-m, -m_{0}}_{\rho}$ such that 
$1 - f\ast g_{\theta}' \in  {\cal S}({\mathbb R}^{d})$ then $g_{\theta}-g'_{\theta} \in {\cal S}({\mathbb R}^{d})$ 
and $1 - g_{\theta}'\ast f \in  {\cal S}({\mathbb R}^{d})$. Thus parametrix is unique up to an addition of element from 
 the Schwartz space.

The above definition of classes 
$\Gamma^{m}_{\rho}$, $\Sigma_{\rho}^{m}$ and ${\rm H}\Gamma^{m, m_{0}}_{\rho}$ 
can be extended in a straightforward way to matrix-valued functions. 
The absolute value $|f(x)|$ in formulas (\ref{asymptotics}), (\ref{sigma_space}) in that case should be understood 
as a matrix norm. Matrix-valued functions belonging to the class 
$\Gamma^{m}_{\rho}$ 
are same as
matrices whose
entries are functions from this class. (This is wrong however for the class  
${\rm H}\Gamma^{m, m_{0}}_{\rho}$.) 
 The pointwise and Moyal multiplications should now be combined with matrix multiplication.

If $\theta$ is nondegenerate we can consider elements from ${\rm H}\Gamma^{m, m_{0}}_{\rho}$ as pseudodifferential operators. 
We will denote this set of operators by ${\rm H}\Gamma^{m, m_{0}}_{\rho, \theta}$ reserving the notation 
 ${\rm H}\Gamma^{m, m_{0}}_{\rho}$ for their symbols.
Such operators are called hypoelliptic (elliptic in the case $m=m_{0}$). These operators act on 
 $ {\cal S}({\mathbb R}^{d/2})$ and can be considered there as Fredholm operators. That is   kernels of  
such an operator $A$ and of its adjoint $A^{\dagger}$ are finite-dimensional and the difference 
${\rm ind}(A) = {\rm dim}({\rm Ker} A) - {\rm dim}({\rm Ker} A^{\dagger})$ is called an index of operator $A$. 
This index can be calculated in the following way. Assume $A$ is a hypoelliptic operator 
corresponding to an $n\times n$ matrix valued function $a(x)\in {\rm H}\Gamma^{m, m_{0}}_{\rho}$.  
Since the matrix $a(x)$  is invertible for large $|x|$ we have a mapping from large   sphere 
$S^{d-1}$ into $GL(n)$. More precisely $a(x)$ specifies a homotopy class of  mappings $S^{d-1}\to GL(n)$.
It is known that for $n\ge d/2$ the homotopy group $\pi_{d-1}(GL(n))$ is isomorphic to integers $\mathbb Z$.     
In particular for $d=4$ the homotopy class of a map from $S^{3}$ to $GL(n)$ is characterized by an integer 
$$
k \in \pi_{3}(GL(n)) \cong \pi_{3}(U(n)) \cong {\mathbb Z} \, .   
$$
The integer $k$ coincides (up to a sign) with the index of operator $A$. 
The relative sign factor is given by the sign of the Pfaffian: ${\rm sign}({\rm Pfaff}(\theta ))$.
 It comes from the operator - function correspondence set by 
formula (\ref{Moyal2}). The easiest way to see this effect is by noticing that $\theta_{i}\mapsto -\theta_{i}$ exchanges the 
creation and annihilation operators (c.f. formula (\ref{z_vs_a}). 
A general formula for $k$ valid 
for any even $d$ looks as 
follows (see for example \cite{Hermander})
\begin{equation}\label{index}
k = {\rm ind}(A) = -{\rm sign}({\rm Pfaff}(\theta ))
\left( \frac{i}{2\pi}\right)^{d/2}\frac{(d/2-1)!}{(d-1)!}\int_{S^{d-1}}{\rm tr}(a^{-1}da)^{d-1}  
\end{equation}
We readily see from this formula  that while the sign of the integral in (\ref{index}) 
depends on the choice of orientation in ${\mathbb R}^{d}$, the index of operator $A$ does not.

\noindent {\bf Example 1.} Consider a function $f=z$ - a complex coordinate on ${\mathbb R}^{2}$ 
introduced in section \ref{conn}. Evidently 
it belongs to the space  ${\rm H}\Gamma^{1, 1}_{1}$. The corresponding operator is 
 $\sqrt{\theta}a$ where $a$ is the annihilation operator. Its index is equal to one.

\noindent {\bf Example 2.} Consider a matrix valued function 
$$
D = \left( \begin{array}{cc} 
 z_{2} & - \bar z_{1} \\
- z_{1} & -\bar z_{2} 
\end{array} \right)
$$ 
where $z_{\alpha}$ are complex coordinates (\ref{cc}) on ${\mathbb R}^{4}$.
It is clear that the symbol $D$ belongs to ${\rm H}\Gamma^{1,
1}_{1}$. 
Note that if one adds  a constant matrix to $D$ the resulting function stays in 
 ${\rm H}\Gamma^{1, 1}_{1}$ (lower order terms in $z$ do not spoil ellipticity). 
To obtain the corresponding elliptic 
operator we replace $z_{\alpha}$, $\bar z_{\alpha}$ by  operators 
proportional to creation or annihilation operators according to (\ref{z_vs_a}).
Taking for example $\theta_{1}, \theta_{2}>0$ we get an operator 
$$
\hat D = \left( \begin{array}{cc} 
 \sqrt{\theta_{2}}a_{2} & -\sqrt{\theta_{2}}a_{1}^{\dagger} \\
-\sqrt{\theta_{2}} a_{1} & -\sqrt{\theta_{2}} a_{2}^{\dagger} 
\end{array} \right) \, . 
$$
It is easy to see that this operator has a one-dimensional kernel spanned by the vector 
$$
\left( \begin{array}{c}
|0\rangle \\
0 \end{array} \right)
$$ 
where $|0\rangle$ is the Fock space vacuum. The adjoint operator
${\hat D}^{\dagger}$ does not have zero modes. 
Thus ${\rm ind}(\hat D)=1$.

A more general construction of an element of  ${\rm H}\Gamma^{1, 1}_{1}$ that works in  any even dimension is 
\begin{equation} \label{D2}
D = \sum_{i = 1}^{d} \gamma_{i}x^{i}
\end{equation}
where $\gamma_{i}$ are Dirac gamma-matrices. The inclusion $D\in {\rm H}\Gamma^{1, 1}_{1}$ immediately follows 
from the relation $D^{2} = {\bf x}\cdot {\bf x}$. One can also show that ${\rm ind}(\hat D) = 1$.

Let us assume that a pseudodifferential operator $A\in {\rm H}\Gamma^{m, m_{0}}_{\rho, \theta}$ 
has no zero modes, i.e., ${\rm Ker} A=0$.  Then we can construct an operator $T = A(A^{\dagger}A)^{-1/2}$ obeying 
$T^{\dagger}T=1$. It is easy to check that both $T$ and $T^{\dagger}$ belong to $ {\rm H}\Gamma^{0, 0}_{\rho, \theta}$. 
Since the  parametrix  of any operator is uniquely defined  up to an operator with Schwartz class matrix elements the 
operator  
$T^{\dagger}$ is a parametrix of $T$ and $1-TT^{\dagger}$ has matrix elements from  $ {\cal S}({\mathbb R}^{d})$ 
(i.e. belongs to ${\cal S}({\mathbb R}_{\theta}^{d})$). 
 We can  apply this construction to the operator $\hat D$ considered above. In that case we have 
$T = \hat D(\hat D^{\dagger}\hat D)^{-1/2}\in {\rm H}\Gamma^{0, 0}_{1, \theta}$. This operator has index 1.

We can use such an operator $T$ to give a rigorous proof of the isomorphism of $\cal A$-modules 
${\cal F}^{k}\oplus {\cal A}^{N}$ and ${\cal A}^{N}$ where $\cal A$ is any of the algebras introduced 
in section \ref{algebras}. The construction of the isomorphism goes as follows. Take any operator $T$ 
satisfying $T^{\dagger}T=1$ and having index ${\rm ind}(T)=k$. For instance we can take  
$T = (\hat D(\hat D^{\dagger}\hat D)^{-1/2})^{k}$ constructed with a help of the matrix valued function $D$ given by 
(\ref{D2}). Since $T^{\dagger}T=1$ and the index of $T$ equals  $k$ we have 
${\rm dim}(T^{\dagger})=k$. Consider a module $E={\rm Ker} T^{\dagger}\oplus {\cal A}^{N}$ and a mapping 
$M:E \to {\cal A}^{N}$ acting as $(\xi, x)\mapsto \xi + Tx$. The inverse map transforms 
$y\in  {\cal A}^{N}$ into $(\Pi y, T^{\dagger}y)$. It remains to prove that ${\rm Ker} T^{\dagger}$ is isomorphic 
to ${\cal F}^{k}$. Note that the map $\Pi=1-TT^{\dagger}$ is a projector of  ${\cal A}^{N}$ onto 
${\rm Ker} T^{\dagger}$. By construction $\Pi$ belongs to ${\cal S}({\mathbb R}_{\theta}^{d})$.
 From this point on
the proof of the isomorphism  ${\rm Ker} T^{\dagger}\cong {\cal F}^{k} $ is essentially a generalization of our 
argument from section \ref{modules} showing that $\cal F$ is projective. Namely one can always choose  a basis in our 
Hilbert space in such a way that
the matrix elements of $\Pi$ are of the form  $\langle x|\Pi|x'\rangle = \sum_{i=1}^{k} \bar \alpha_{i}(x)\alpha_{i}(x')$ 
where the functions $\alpha_{i}(x)$ belong to ${\cal S}({\mathbb R}^{d/2})$ and are orthogonal to each other.  
An arbitrary element of ${\rm Ker} T^{\dagger}$ can be written as $\Pi\cdot \hat a$ where $\hat a \in {\cal A}^{N}$. 
It has matrix elements $\langle x|\Pi \hat a|x'\rangle = \sum_{i=1}^{k} \bar \alpha_{i}(x)\rho_{i}(x')$ where 
$\rho_{i}(x)\in {\cal S}({\mathbb R}^{d/2})$. The mapping $\Pi\cdot \hat a \mapsto (\rho_{i}(x))\in {\cal F}^{k}$ 
establishes the desired isomorphism.

It is easy to show that by means of the above isomorphism the trivial connection $\partial_{i}$ on  
${\cal A}^{N}$ induces a connection $\Pi\partial_{i}\Pi$ on ${\rm Ker} T^{\dagger}$. Let us calculate a 
connection $D_{\alpha}$ on ${\cal A}^{N}$ that corresponds to the connection 
$$
\left( \begin{array}{cc}
\Pi\partial_{i}\Pi & 0\\
0& \partial_{i} 
\end{array}
\right)
$$
on ${\rm Ker} T^{\dagger}\oplus {\cal A}^{N}$. We have a sequence of maps 
$$
{\cal A}^{N}\ni y \mapsto (\Pi y , T^{\dagger}y) \mapsto (\Pi \partial_{i}\Pi y, \partial_{i}T^{\dagger}y) 
\mapsto \Pi \partial_{i}\Pi y + T\partial_{i}T^{\dagger}y \, . 
$$
Therefore $D_{i} =  \Pi \partial_{i}\Pi + T\partial_{i}T^{\dagger}$.

More generally we can consider a connection 
\begin{equation} \label{gen_con}
\left( \begin{array}{cc}
\Pi\partial_{i}\Pi & 0\\
0& \partial_{i} 
\end{array}
\right)   +
 \left( \begin{array}{cc} 
B_{i} & \langle I_{i}| \\
|J_{i}\rangle & \hat R_{i}  
\end{array} \right)
\end{equation}
on ${\rm Ker} T^{\dagger}\oplus {\cal A}^{N}$. Under the isomorphism $M$ it goes into a connection 
$D_{i} + A_{i}$ where 
\begin{equation} \label{new_con}
A_{i} = \Pi B_{i} \Pi + \Pi |I_{j}\rangle\langle I_{j}|T^{\dagger} + T |J_{j}\rangle\langle J_{j}|\Pi + T\hat R_{j} T^{\dagger}
\end{equation}
or, less invariantly but more explicitly,  assuming  $\Pi = \sum_{\alpha=1}^{k} |\alpha \rangle\langle \alpha |$ we can write 
$$
A_{i} = |\alpha\rangle B^{\alpha \beta}_{i}\langle \beta | + |\alpha\rangle\langle I^{\alpha}_{i} |T^{\dagger} 
+ T|J_{\alpha i}\rangle\langle \alpha| + T\hat R_{i} T^{\dagger} \, . 
$$
Identifying ${\rm Ker} T^{\dagger}$ with ${\cal F}^{k}$ we obtain a correspondence between connections on 
${\cal F}^{k}\oplus  {\cal A}^{N}$ and connections on $ {\cal A}^{N}$.
We can rewrite the above connection  $D_{i} + A_{i}$ in the form $ T\partial_{i}T^{\dagger} + \Pi\partial_{i}\Pi + \nu_{i}$. 

Furthermore if we assume that both operators $T$ and $T^{\dagger}$ belong 
to ${\rm H}\Gamma^{0,0}_{1, \theta}$ then one can  argue that $\nu_{i}$ has the same behavior at infinity as the block $\hat R_{i}$. 
This follows from formula (\ref{new_con}) once we notice that the terms containing $\Pi$ belong to 
${\cal S}({\mathbb R}_{\theta}^{d})$ and multiplying by $T$ or $T^{\dagger}$ does not change the asymptotic behavior.  
In particular (under the above assumptions on $T$ and $T^{\dagger}$)  
$\nu_{i} \in \Gamma^{m}(R_{\theta}^{d})$ if and only if $\hat R_{i}$ belongs to the same algebra. 

{\it We shall call a connection (gauge field) (\ref{gen_con}) on 
a $\Gamma_{\theta}$-module ${\cal F}_{k, N}$ whose lower right block $\hat R_{i}$ belongs 
to  $\Gamma_{\theta}$ a $\tilde \Gamma_{\theta}$-connection (gauge field). } As the name reveals these connections 
are precisely those connections on a 
given $\Gamma_{\theta}$-module that can be extended to the corresponding $\tilde \Gamma_{\theta}$-module. (As we already discussed 
once we add the unit element the  endomorphisms of $\tilde {\cal A}^{N}$ 
correspond to matrices with entries from unitized algebra.)

\subsection{ Fields gauge trivial at infinity.} \label{gauge_triv}
In this section we will analyze instantons on noncommutative ${\mathbb R}^{4}$ considering them as gauge fields  
on ${\Gamma_{\theta}}^{N}$. As usual we define an instanton as a gauge field 
satisfying (\ref{inst_hol}), (\ref{inst_real}). However we should also impose additional constraints on our gauge fields 
that will guaranty that the corresponding Euclidean Yang-Mills action is finite. We will require that (anti)instantons are 
gauge trivial at infinity in the following sense. 

We first give a definition of a connection gauge trivial at infinity that works for arbitrary $\theta$, in particular 
for a degenerate one. 
Consider an element  $T\in Mat_{N}({\cal A}) $ where $\cal A$ is one of the algebras ${\mathbb R}_{\theta}^{d}$, 
$\Gamma_{\theta} $, ${\cal S}({\mathbb R}_{\theta}^{d}) $. 
Assume also that $T\in {\rm }{\rm H}\Gamma^{m, m_{0}}_{1 }$. The 
parametrix $S$ of $T$ is represented by a $N\times N$  
  matrix belonging to     
${\rm H}\Gamma^{-m, -m_{0}}_{1}$ and we have elements  $\Pi = 1-T\ast S$, $\Pi'=1-S\ast T$ 
belonging to $Mat_{N}({\cal S}({\mathbb R}_{\theta}^{d}))$. 
Then we say that  a connection $\nabla_{i}: {\cal A}^{N} \to {\cal A}^{N}$ is gauge trivial at infinity if 
it can be written in a form 
\begin{equation}
\nabla_{i} = T\cdot \partial_{i}\cdot S + \Pi\cdot \partial_{i}  + \nu_{i}
\end{equation}
for some operators $T$ and $S$ satisfying the assumptions above and some endomorphism $\nu_{i}\in \Gamma_{\theta} $. 
In this formula $T$ and $S$ stand for operators acting on $ {\cal A}^{N}$ by  a star-multiplication from the right (combined 
with the usual matrix multiplication). 
The term $\Pi\cdot \partial_{i}$ is needed here to ensure that operator $\nabla_{i}$ is indeed a connection. 
This becomes manifest when we rewrite the above expression as 
$ \nabla_{i} = \partial_{i} + T[\partial_{i}, S] + \nu_{i}$ that has the form $\partial_{i} + \mbox{endomorphism}$.  
Note also that in the commutative case $\theta=0$ the above definition coincides with the standard definition of 
gauge triviality at infinity.

A gauge field, as we defined it, is a unitary (antihermitian) connection. Assume further that the matrix $\theta$ is 
nondegenerate. This allows us to consider $T$ and $S$  as pseudo-differential operators. 
To ensure unitarity we set in the above definition $S=T^{\dagger}$. Then automatically 
$T, T^{\dagger}\in {\rm H}\Gamma^{0,0}_{1, \theta}$. 
Furthermore
 in the above  definition we  have a freedom of shifting operators $T$ and $S$ by operators from  
${\cal S}({\mathbb R}_{\theta}^{d})$. 
We can use that freedom to set either $\Pi'$ or $\Pi$, depending on ${\rm ind}(T)$, to zero. 
If we can set  $\Pi=0$ then $\Pi'$ is a projector and vice versa.  {\it  
We say that $\nabla_{i}$ is a gauge field 
 gauge trivial at infinity if it can be represented in the form  
\begin{equation} \label{g_triv}
\nabla_{i} = T\cdot \partial_{i}\cdot T^{\dagger} + \Pi\cdot \partial_{i} \cdot \Pi  + \nu_{i}
\end{equation}
where $\nu_{i}$ is an antihermitian endomorphism from $\Gamma_{\theta} $ and  $T, T^{\dagger}\in  {\rm H}\Gamma^{0,0}_{1, \theta}$. 
The space of all such 
connections has connected components  labeled by the index $I$ of operator $T$.     } 
We fix one operator $T=T_{I}$ for every value of the index $I$ and denote by ${\cal G}_{I}$ the set 
of gauge fields having the form (\ref{g_triv}) with the operator $T=T_{I}$.
Specializing further to the case of (anti)instantons we will assume from now on  that $d=4$, ${\cal A}=\Gamma_{\theta}$.

As we explained in the previous section if operator $T$ has a negative index we may use a pair $T$, $T^{\dagger}$
to establish an isomorphism 
of  a general  module 
${\cal F}_{k,N}={\cal F}^{k}\oplus {\Gamma_{\theta}}^{N}$  
with $ {\Gamma_{\theta}}^{N}$. 
The results of the previous section imply   that {\it $\tilde \Gamma_{\theta}$ gauge fields on a 
$\Gamma_{\theta}$-module ${\cal F}^{k}\oplus { \Gamma_{\theta}}^{N}$ are in one 
to one correspondence with a class of gauge fields   ${\cal G}_{k}$  gauge trivial at infinity 
on a module ${ \Gamma_{\theta}}^{N}$.}  
Moreover it is easy to check that the condition of gauge triviality at infinity 
 implies the finiteness of the Euclidean action (it follows  essentially 
from   inclusions $\Gamma^{m}({\mathbb R}_{\theta}^{4})\cdot \Gamma^{n}({\mathbb R}_{\theta}^{4})
\subset \Gamma^{m+n}({\mathbb R}_{\theta}^{4})$).  
Notice that   a connection 
on ${\cal F}^{k}\oplus \Gamma_{\theta}^{N}$  can be extended to a connection on a $\tilde \Gamma_{\theta}$-module 
 ${\cal F}^{k}\oplus \tilde \Gamma_{\theta}^{N}$. The former fact is implied from the inclusion $\nu_{i} \in \Gamma_{\theta}$.
This remark provides a bridge between the considerations of the present section and those of section \ref{inst1}.


\subsection{ Noncommutative ADHM construction} \label{ADHM22}
Now we are fully armed  to discuss the noncommutative ADHM construction in detail.
In part I of this review we explained how one can construct a connection on a  module $E$ specified by 
a projector $P$ acting in a free module ${\cal A}^{N}$ from any   connection $\nabla^{0}_{i}$ on $ {\cal A}^{N}$. 
Namely  the operator $\nabla = P\cdot \nabla^{0}_{i}\cdot P$ descends to a connection on $E$. 
In  particular one may take $\nabla_{i}^{0}$ to be just the trivial connection $\partial_{i}$. This gives rise to the so called 
Levi-Civita connection. The ADHM construction 
of (anti)self-dual connections starts with solutions to some matrix equations that are further used to construct 
 such a projector $P$ 
that  the corresponding Levi-Civita connection $P\cdot \partial_{i}\cdot P$ is (anti)self-dual. The precise construction 
goes as follows. 

For the discussion in this subsection we will assume unless the nondegeneracy is explicitly stated  that $\theta$ is arbitrary. 
Let $V$ and $W$ be a pair of complex vector spaces of dimensions $k$ and $N$ respectively. 
Consider a $\Gamma_{\theta}$-linear operator 
$$
D^{\dagger}: (V\oplus V\oplus W)\otimes { \Gamma_{\theta}} \to (V\oplus V)\otimes {\Gamma_{\theta}}
$$ 
defined by the formula 
\begin{equation} \label{DD}
D^{\dagger} = \left( \begin{array}{ccc} 
-B_{2} + \hat z_{2} & B_{1} - \hat z_{1} & I \\
B_{1}^{\dagger} -\widehat{\bar z_{1}}&B_{2}^{\dagger} - \widehat{\bar z_{2}}& J^{\dagger} \end{array} \right)
\end{equation}
where $B_{1}, B_{2}:V\to V$, $I:W\to V$, $J:V\to W$  are linear mappings, 
 $\hat z_{\alpha}$ $\widehat{\bar z_{1}}$ are operators acting by star multiplication by 
the corresponding  complex coordinates (\ref{complex_str}). 
The requirement that $D^{\dagger}$ satisfies  
$$
D^{\dagger}D = \left( \begin{array}{cc} \Delta & 0\\
0&\Delta 
\end{array} \right)
$$
where $\Delta$ is some  operator $\Delta: V\otimes \Gamma_{\theta} \to V\otimes \Gamma_{\theta}$  
is equivalent to  ADHM equations (\ref{ADHM1}), (\ref{ADHM2}) with 
$\zeta_{r} = -(\theta_{1} + \theta_{2})$.
More precisely the vanishing of the off-diagonal blocks in $D^{\dagger}D$ is equivalent to the first ADHM equation 
 (\ref{ADHM1}) and the equality of the diagonal blocks is equivalent to the second one  (\ref{ADHM2}).
 The difference between the present $\zeta_{r}$ and the one 
that appeared in section \ref{inst1} can be absorbed into redefinition of matrices $B_{1}$, $B_{2}$, $I$, $J$. 
It is easy to compute that 
\begin{eqnarray} \label{delta}
\Delta &=& (B_{1} - \hat z_{1} )(B_{1}^{\dagger} -  \widehat{\bar z_{1}}) + 
(B_{2} - \hat z_{2})(B_{2}^{\dagger} -  \widehat{\bar z_{2}}) + II^{\dagger} = \nonumber \\
&& (B_{1}^{\dagger}-  \widehat{\bar z_{1}})(B_{1}-\hat z_{1}) + (B_{2}^{\dagger} -  \widehat{\bar z_{2}})
(B_{2} - \hat z_{2}) +J^{\dagger}J \, . 
\end{eqnarray}

Let us now show that provided $\theta^{ij}$ is nondegenerate the 
operator $\Delta  $ does not have any zero modes. Recall that our algebra $\Gamma_{\theta}$ has a natural representation 
in terms of integral operators in ${\cal S}({\mathbb R}^{2})$, i.e. the Fock module $\cal F$.
If there exists an operator  
$\hat \Phi \in V\otimes \Gamma_{\theta}$ satisfying $\Delta\cdot \hat \Phi = 0$ then any vector $|\phi\rangle \in {\cal F}$ 
 of the form $|\phi\rangle = \hat \Phi |\phi \rangle$ is annihilated by $\Delta$. Using the first representation in (\ref{delta}) 
we obtain an equation 
$$
 \langle \phi|\Delta |\phi \rangle = \| (B_{1}^{\dagger} -  \widehat{\bar z_{1}})\phi \|^{2} + 
\| (B_{2}^{\dagger} -  \widehat{\bar z_{2}})\phi\|^{2} + \| I^{\dagger}\phi\|^{2}=0
$$  
that implies 
$$
(B_{1}^{\dagger} -  \widehat{\bar z_{1}})|\phi\rangle =0 \, , \quad (B_{2}^{\dagger} -  \widehat{\bar z_{2}})|\phi\rangle=0 \, , \quad 
I^{\dagger}|\phi\rangle = 0 \, . 
$$
Analogously from the second representation of $\Delta$ in  (\ref{delta}) we obtain 
$$
(B_{1}-\hat z_{1})|\phi \rangle  =0 \, , \quad (B_{2} - \hat z_{2})|\phi \rangle =0 \, , \quad J |\phi \rangle =0 \, . 
$$
The set of operators $\hat z_{\alpha}$, $\widehat{\bar z_{\alpha}}$ can be identified 
 up to numerical factors $|\theta_{\alpha}|^{-1/2}$ with the set of creation and 
annihilation operators $a^{\dagger}_{\alpha}$, $a_{\alpha}$.  The particular identifications depend on the signs of 
eigenvalues $\theta_{\alpha}$  according  to (\ref{z_vs_a}). The statement $\phi = 0$ then follows from the fact that 
the equations 
$$
(B + a^{\dagger}_{1})|\phi \rangle = 0 \, , \qquad  (B' + a^{\dagger}_{2})|\phi \rangle = 0
$$
have only trivial solution.

Thus when $\theta$ is nondegenerate the operator $\Delta$ is invertible. The condition of nondegeneracy can be 
weakened. The proof above can be easily modified for the assumption  that $\theta^{ij}$ is merely nonzero 
(see  \cite{revisited} for details).  
 In the commutative  $\theta^{ij}=0$  case the nondegeneracy of  $\Delta$  is a separate assumption. 
Assuming the nondegeneracy of $\Delta$   we can split the module $(V\oplus V\oplus W)\otimes \Gamma_{\theta}$ into 
a direct sum of two subspaces ${\cal E}\oplus {\cal E}'$ where 
 ${\cal E}={\rm Ker}(D^{\dagger})$  and  ${\cal E}'=D(V\oplus V\oplus W)\otimes \Gamma_{\theta}\equiv {\rm Im}(D)$.  
These  subspaces  are in fact submodules because $D$ and $D^{\dagger}$ are $\Gamma_{\theta}$-linear operators.
 Each of these submodules can be identified with images of orthogonal projectors 
${\cal E} = \Pi(V\oplus V\oplus W)\otimes \Gamma_{\theta}$, ${\cal E}' = (1 - \Pi )(V\oplus V\oplus W)\otimes \Gamma_{\theta}$ where 
\begin{equation} \label{Pi}
\Pi = 1 - D\cdot (D^{\dagger} D)^{-1}\cdot D^{\dagger} \, . 
\end{equation} 
The inverse of operator $ D^{\dagger} D$ exists under the assumption  that the operator $\Delta$ is invertible 
(that as we proved above is always true for nondegenerate $\theta^{ij}$ and should be taken as an additional requirement 
otherwise). 

In summary starting with a solution to the noncommutative ADHM equations we constructed an orthogonal projector 
$\Pi$. The claim then is that the Levi-Civita connection $\Pi \partial_{i} \Pi$ induced on the  module   
$\cal E$ satisfies the self-duality equation. Let us give an explicit proof of it. 
The curvature of connection $\Pi \partial_{i} \Pi$ is a two-form that restricted on ${\cal E}$ can be written as 
\begin{equation} \label{11}
F = d\Pi\wedge d\Pi = d( D (D^{\dagger} D)^{-1} D^{\dagger})\wedge  d( D (D^{\dagger} D)^{-1} D^{\dagger}) \, .
\end{equation}
Using 
$$
d  (D^{\dagger} D)^{-1} = - (D^{\dagger} D)^{-1}d(D^{\dagger}D)(D^{\dagger} D)^{-1}
$$
we obtain 
$$ 
d( D (D^{\dagger} D)^{-1} D^{\dagger}) = \Pi d(D)(D^{\dagger} D)^{-1}D^{\dagger} + D(D^{\dagger} D)^{-1}d(D^{\dagger}) \Pi \, . 
$$
Substituting this expression into (\ref{11}) we get 
$$
F = \Pi  d(D)(D^{\dagger} D)^{-1} \wedge d(D^{\dagger})\Pi + D(D^{\dagger} D)^{-1}d(D^{\dagger})\Pi \wedge d(D)
(D^{\dagger} D)^{-1}D^{\dagger} \, . 
$$
The second term in this expression vanishes when restricted to $\cal E$. 
We can now plug the concrete expressions for operators $D$ and $D^{\dagger}$ corresponding to (\ref{DD}) 
 into the second term. Using the identity  
$$
\left( \begin{array}{cc} d\bar z_{2} & -dz_{1} \\
-d\bar z_{1} & -dz_{2} \end{array} \right)\wedge  \left( 
 \begin{array}{cc} d z_{2} & -dz_{1} \\
-d\bar z_{1} & -d\bar z_{2} \end{array} \right) = 2\left( \begin{array}{cc} if_{1} & f_{2} +if_{3} \\
f_{2}-if_{3} & -if_{1} \end{array} \right) 
$$
where 
$$
f_{1} = dx^{3}\wedge dx^{1} + dx^{2}\wedge dx^{4} \, , \quad f_{2} = dx^{1}\wedge dx^{2} + dx^{3}\wedge dx^{4} \, , 
\quad f_{3} = dx^{4}\wedge dx^{1} + dx^{3}\wedge dx^{2}
$$
are basic self-dual forms,  
 we obtain for the curvature
$$
F = 2\Pi\left( \begin{array}{ccc} i\Delta^{-1} f_{1} & \Delta^{-1}( f_{2} +if_{3}) & 0 \\
\Delta^{-1}(f_{2}-if_{3}) & -i\Delta^{-1}f_{1} & 0 \\
0&0&0 \end{array}\right) \Pi 
$$
that is evidently  self-dual. 

The above construction can be represented in a slightly different form. Instead of singling out the  
submodule ${\cal E}$ by the condition ${\cal E}={\rm Ker}D^{\dagger}$ we can take as a starting point 
an isometric $\Gamma_{\theta}$-linear map 
\begin{equation}\label{omega}
 \omega: \tilde {\cal E} \to (V\oplus V\oplus W)\otimes {\Gamma_{\theta}}
\end{equation}
that maps some $\Gamma_{\theta}$-module $ \tilde {\cal E}$ onto ${\cal E}$. 
For the mapping $\omega$ to be 
an isometry one has to have $\omega^{\dagger}\omega=1$.
This mapping induces a self-dual connection on $\cal E$ that can be written as 
 $\omega^{\dagger}\cdot \partial_{\alpha}\cdot \omega$. 

\subsection{ADHM instantons as gauge fields trivial at infinity and  ${\cal F}_{k, N}$-modules.} \label{inst2}
In section \ref{gauge_triv} we introduced an important notion of gauge fields trivial at infinity. 
Such fields have a finite Euclidean action. Here we would like to show that the instanton solutions constructed 
via the noncommutative ADHM construction are  gauge 
trivial at infinity. 

To this end we first construct a module $\tilde {\cal E}$ that is isometrically isomorphic to ${\rm Ker}D^{\dagger}$ by means 
of some embedding $\omega$ (\ref{omega}). Throughout the discussion in this subsection we assume $\theta^{ij}$ to be nondegenerate.
An element  $\psi\in {\rm Ker}D^{\dagger}$ 
can be represented by a column  vector 
$$
\psi = \left( \begin{array}{c} 
\phi\\ 
\zeta
\end{array} \right)  = \left( \begin{array}{c} 
\phi_{1}\\
\phi_{2}\\ 
\zeta
\end{array} \right)
$$
where $\phi_{i}\in V\otimes \Gamma_{\theta}$ and  $\zeta \in W\otimes\Gamma_{\theta}$.
The equation $D^{\dagger}\psi = 0$ then reads 
\begin{equation} \label{zero}
A\phi + C\zeta = 0 
\end{equation} 
where 
$$
A\phi =  \left( \begin{array}{cc} 
-B_{2} + \hat z_{2} & B_{1} - \hat z_{1}  \\
B_{1}^{\dagger} -\widehat{\bar z_{1}}&B_{2}^{\dagger} - \widehat{\bar z_{2}} \end{array} \right)
 \left( \begin{array}{c} 
\phi_{1}\\
\phi_{2}
\end{array} \right)
\, ,   \qquad  C\zeta = \left( 
\begin{array}{c} I\zeta \\ 
J^{\dagger}\zeta \end{array} \right) \, . 
$$

The above operator $A: (V\oplus V)\otimes \Gamma_{\theta}\to  (V\oplus V)\otimes \Gamma_{\theta}$ is defined as 
an operator of left star multiplication by the corresponding matrix-valued function.  
This function  belongs  to the class  ${\rm H}\Gamma^{1,1}_{1}$. The corresponding  elliptic pseudo-differential operator 
(not to be confused with the operator $A$ itself) 
can be expressed via creation and annihilation operators according to (\ref{z_vs_a}). 
 It is not hard to see from those expressions 
that the index of $A$ equals ${\rm dim}V$ when ${\rm Pfaff}(\theta )=\theta_{1}\theta_{2}>0$ 
and ${\rm Ind}(A)=- {\rm dim}V$ when $\theta_{1}\theta_{2}<0$. We will consider these two cases separately 
beginning with the $\theta_{1}\theta_{2}>0$ case.  
In this case the operator $A$ has a parametrix $Q$ that 
without loss of generality can be assumed to satisfy  $AQ=1$ and $QA=1-P$ where $P\in {\cal S}({\mathbb R}^{4}_{\theta})$. 
  By multiplying equation (\ref{zero}) by $Q$ from 
the left we obtain that $\phi = \phi_{0} - QC\zeta$ where $\phi_{0}=P\phi$ belongs to the kernel 
${\rm Ker}A\cong {\cal F}^{k}$. 
We take now $\tilde {\cal E}={\cal F}^{k}\oplus (W\otimes \Gamma_{\theta})$  where we identify ${\cal F}^{k}$ with 
a submodule ${\rm Ker}A\subset (V\oplus V)\otimes \Gamma_{\theta}$. 
Consider a mapping 
$\nu : \tilde {\cal E}\to (V\oplus V\oplus W)\otimes \Gamma_{\theta}$ acting as 
\begin{equation}\label{nu}
 \left( \begin{array}{c} \rho \\
 \xi \end{array} \right) \mapsto 
 \left( \begin{array}{c} \rho - QC\xi \\
 \xi \end{array} \right) 
\end{equation}
where $\rho \in {\rm Ker}A\subset (V\oplus V)\otimes \Gamma_{\theta}$, $\xi W\otimes \Gamma_{\theta}$. This mapping is evidently 
$\Gamma_{\theta}$-linear and has trivial kernel. To check that the image of this map belongs to $\cal E$ we first note 
that as it follows from $AQ=1$    a point in the image of $\nu$: $\phi=  \rho - QC\xi$, 
$\zeta=\xi$ satisfies equation (\ref{zero}). Next,  as we have 
 $Q\in {\rm H}\Gamma^{-1,-1}_{1, \theta}$ and $C\in {\rm H}\Gamma^{0,0}_{1, \theta}$, 
the element  $(\rho - QC\xi, \xi)  $ belongs to $(V\oplus V\oplus W)\otimes {\Gamma_{\theta}}$.
 It is also easy to check that the inverse mapping is given by 
the formula 
$$
\nu^{-1} : \left( \begin{array}{c} u\\
 \xi \end{array} \right) \mapsto 
 \left( \begin{array}{c} u+  QC\xi \\
 \xi \end{array} \right)
$$ 
and it maps $\cal E$ into $\tilde {\cal E}$. It remains to modify the isomorphism $\nu$ to make it an 
isometry. This can be done by taking $\omega = \nu(\nu^{\dagger}\nu)^{-1/2}$. 
It is not hard to derive from the explicit form (\ref{nu}) that $\omega$ and $\nu$ have the same asymptotic behavior 
 at infinity. We imply from this that  the $W\otimes \Gamma_{\theta} \to W\otimes \Gamma_{\theta}$  block of the connection 
$\omega^{\dagger}\partial_{\alpha}\omega: \tilde {\cal E}\to \tilde {\cal E} $
 has the from $\partial_{\alpha} + a[\partial_{\alpha}, b]$ where 
$a$ and $b$ are some operators having the form $a=1 + a'$, $b=1 +b'$ where 
$a', b'\in \Gamma^{-1}({\mathbb R}_{\theta}^{4})$. 
 Therefore $\omega^{\dagger}\partial_{\alpha}\omega$ 
is a $\tilde \Gamma_{\theta}$-connection and using the correspondence explained in section \ref{math} we conclude that 
(for $\theta_{1}\theta_{2}>0$) the noncommutative ADHM (anti)instanton solution is indeed gauge trivial at infinity.

For simplicity we used above the assumption that the parametrix $Q$ satisfies  $AQ=1$ and $QA=1-P$. The general  
situation   when one has $AQ=1-P'$,  $QA=1-P$, $P, P'\in {\cal S}(R_{\theta}^{4})$ can be reduced 
to this particular case by a change of variables: $ \phi' = \phi$, $\zeta'=\zeta + M\phi$ where $M$ can be  chosen 
in such a way that the above arguments will go through when applied to  the equation $(A-CM)\phi' + C\zeta = 0$.

Let us consider now the case $\theta_{1}\theta_{2}<0$. 
In this case the matrix-valued function representing the operator  $A$
above gives rise to an elliptic 
operator of index $-{\rm dim}V$. Then one can assume that there exists a parametrix 
$Q:(V\oplus V)\otimes \Gamma_{\theta}\to  (V\oplus V)\otimes \Gamma_{\theta}$   satisfying $QA=1$, $AQ=1-P'$ where 
$P'\in  {\cal S}(R_{\theta}^{4})$ is a projector.  Multiplying equation (\ref{zero}) describing the kernel of $D^{\dagger}$ 
(\ref{zero}) by operator $Q$ from 
the left we obtain $\phi = -QC\zeta$. Substituting this back into  (\ref{zero}) we get $P'C\zeta =0$. This means that we 
can identify the space of solutions to $D^{\dagger}\psi=0$ with the kernel of operator 
$P'C:W\otimes \Gamma_{\theta} \to (V\oplus V)\otimes \Gamma_{\theta}$. 
Since  $P'$ considered as a pseudo-differential operator in Hilbert space has a finite-dimensional image, i.e., is an  
operator of finite defect, the operator solutions to $P'C\zeta=0$ form a space isomorphic to a module $W\otimes \Gamma_{\theta}$. 
Assume that there is an isometric embedding 
$$
T:W\otimes \Gamma_{\theta} \to W\otimes \Gamma_{\theta}   \, , \quad T\in {\rm H}\Gamma^{0,0}_{1, \theta}
$$ 
such that the image of $T$ coincides with 
 $\tilde {\cal E} = {\rm Ker}P'C$. 
Let us further define a mapping $\lambda: \tilde {\cal E} \to (V\oplus V\oplus W)\otimes \Gamma_{\theta}$ 
as 
\begin{equation}
\lambda (\zeta ) = \left( \begin{array}{c} -QC\zeta\\
 \zeta \end{array} \right) \, . 
\end{equation} 
It is easy to check that this map is an isomorphism and its image coincides with $ {\cal E}$. 
As before one can make this isomorphism an  isometry by modifying 
it as    $\tilde \lambda = \lambda (\lambda^{\dagger}\lambda)^{-1/2}$. Thus ultimately we have a sequence of 
two isometric isomorphisms  
$$
\Gamma_{\theta}^{N}\stackrel{T}{\longrightarrow} \tilde {\cal E}\stackrel{\tilde \lambda}{\longrightarrow} {\cal E} \, . 
$$
The pull-back of the ADHM antiinstanton on the module $\Gamma_{\theta}^{N}$ is a connection 
\begin{equation} \label{pull-back}
\nabla_{\alpha}=(T^{\dagger}\tilde \lambda^{\dagger})\cdot  \partial_{\alpha} \cdot (\tilde \lambda T) 
 :\Gamma_{\theta}^{N}\to \Gamma_{\theta}^{N} \, . 
\end{equation}
 The mapping $\tilde \lambda$ can be explicitly represented in a block form as 
$$
\tilde \lambda = \left( \begin{array}{c} -QC(1 + (QC)^{\dagger}QC)^{-1/2} \\
(1 + (QC)^{\dagger}QC)^{-1/2} \end{array} \right) \equiv 
 \left( \begin{array}{c} \lambda_{1} \\
\lambda_{2}
 \end{array} \right) \, . 
$$
Since  the symbols of pseudodifferential operators corresponding to $QC$ and $ (QC)^{\dagger}$ belong to the class 
${\rm H}\Gamma^{-1,-1}_{1}$ the symbol of $\lambda_{1}$ belongs to the same class and the one  of $\lambda_{2}$ 
has the form $1 + \lambda_{2}'$ where $\lambda_{2}'\in  {\rm H}\Gamma^{-1,-1}_{1}$. 
Substituting this expression into (\ref{pull-back}) we obtain 
$$
\nabla_{\alpha}= T^{\dagger}\partial_{\alpha}T + \lambda_{1}^{*}[\partial_{\alpha}, \lambda_{1}] + 
\lambda_{2}^{*}[\partial_{\alpha}, \lambda_{2}] 
$$
From the asymptotic properties of symbols of $\lambda_{1}$ and $\lambda_{2}$ noted above it follows that 
$\nabla_{\alpha}$ has the form $T^{\dagger}\partial_{\alpha}T + \nu_{i}$ where 
$\nu_{i}\in \Gamma^{-2}({\mathbb R}_{\theta}^{4})\subset \Gamma_{\theta}$. Therefore we see that in 
this form $\nabla_{\alpha}$ has a manifestly gauge trivial at infinity form. 
Note that the assumption on the parametrix $Q$ can be disposed of similarly to  the $\theta_{1}\theta_{2}>0$ case.

Let us comment on the difference in treatments of the $\theta_{1}\theta_{2}>0$ and  $\theta_{1}\theta_{2}<0$ 
cases. In the first case ${\rm Ind}(A)= {\rm dim}V=k$ and with our simplifying assumption on the parametrix the 
pseudodifferential operator corresponding to $A$ has a kernel of dimension $ {\rm dim}V$. Thus $A$ itself 
has a kernel isomorphic to a Fock module ${\cal F}^{k}$ and we took for our module 
$\tilde {\cal E} = {\cal F}^{k}\oplus (W\otimes \Gamma_{\theta})$. In the second case  ${\rm Ind}(A)= -{\rm dim}V=-k$ 
and under our assumptions $A$ has a cokernel isomorphic to ${\cal F}^{k}$. Moreover the construction 
of $\tilde {\cal E}$ that we employed describes $\tilde {\cal E}$ by means of an operator $T$ such that the
pseudodifferential operator corresponding to $T$ has a finite 
defect. For a generic $C$ the dimension  of the defect subspace is precisely $k$ and ${\rm Coker}T$ is isomorphic 
to ${\cal F}^{k}$. Intuitively this means that in this case $\cal E$ is given by means of ``subtraction'' of ${\cal F}^{k}$ 
from the module $W\otimes \Gamma_{\theta}$ in contrast  to adding the same kind of Fock module in the case  $\theta_{1}\theta_{2}>0$.
Another way to describe this situation is based on the consideration of
$\bf Z_2$-graded modules (supermodules and superconnctions). We interpret
"subtraction" of a module as addition of this module with reversed parity.
In this approach one can work with modules over unitized algebra. The phenomenon of "subtraction" of a Fock module 
is also discussed in \cite{Nek_Braden}, \cite{Furuuchi1} from a different point of view.

 One can prove that all  self-dual and antiself-dual gauge fields
that are gauge trivial at infinity can be obtained by means
of ADHM construction \cite{Nikita}.  As in the commutative
situation the proof is based on the consideration of Dirac operator
in the presence of instanton gauge field. The ADHM data can be extracted from the study of
asymptotic behavior of zero modes of this operator.
                                                   
\section{Literature}
 The subject of noncommutative solitons and instantons is currently  an area of active research and a number of new  papers  
grows very fast. 
In this review we covered only  basic facts about these  objects trying to provide a reader with a 
solid mathematical foundation. We refer the reader to  review papers 
\cite{Trieste}, \cite{Komaba}, \cite{DN_rev} for exposition of other topics in noncommutative field theories 
 not covered in the present review. Below we would like to give some rough guide to the existing literature 
that is very subjective and is concentrated on  subjects that were not covered in this review rather than on 
complete historical  references. 

Most of the mathematical results used  in section 2 can be found 
in books \cite{Shubin}, \cite{Rieffel}. The first examples of  exact soliton solutions in Yang-Mills-Higgs type 
systems discussed in sections 
3.1, 3.2 were  considered in \cite{Pol}, \cite{GrNek2}. These examples were  generalized to a  broader set of models in 
 \cite{Unstable}, \cite{Exact}, \cite{GrNek3}. 
Our exposition differs from the approach based on partial isometries that is taken in  most of the papers on the subject 
 and follows the ideas of  \cite{Newappr} that were put forward for noncommutative instantons. 
The stability of the exact solutions in $d=2+1$ dimension is analyzed in \cite{Unstable}. 
Exact  solutions in systems with scalar fields ``in the fundamental representation'' 
were studied in \cite{Terashima_etal}, \cite{Hashimoto}.  

The scalar solitons in noncommutative field theories in 
the limit $\theta \to \infty$ were first studied in paper \cite{NCsol}.  Our exposition of this subject in section 
3.3 is mostly inspired by papers \cite{NCsol}, \cite{Unstable}, \cite{Ganor_etal}, \cite{Gop}. 
The scalar solitons at finite $\theta$ were further studied  in \cite{Zhou}, \cite{Solovyov}, \cite{existence1}
\cite{stab'}. The existence of exact scalar soliton solutions at finite $\theta$ was recently proved in 
\cite{existence_stab}. The questions of stability were addressed in \cite{NCsol}, \cite{stability}, \cite{stab'}
\cite{existence_stab}.

Noncommutative instantons and 
deformation of the  ADHM construction were introduced in paper \cite{NS}. The construction was further refined in 
\cite{Furuuchi1}, \cite{revisited}, \cite{Ho}, \cite{Furuuchi2}, \cite{Furuuchi3}, \cite{Furuuchi4}. 
 Our exposition in section 4 essentially follows 
paper \cite{Newappr}. See also \cite{Trieste} for a complementary review as well as papers \cite{Kapustin_etal}, 
\cite{Tong_etal}, \cite{Rangamani} for related work in  other directions.

The monopole solutions on noncommutative plane are studied in \cite{GrNek1}, \cite{GrNek2}, \cite{GrNek3}.  
A number of other types of solitonic solutions in various noncommutative field theories 
were considered in papers \cite{Pol}, \cite{Jatkar_etal}, \cite{Lee_etal}, \cite{Gorsky_etal}, \cite{Bak}, \cite{Bak_etal}, 
\cite{Bak_Lee}, \cite{Vol_Spr}, \cite{2+1}. 

Both the exact solitonic and the approximate ones at $\theta \to \infty$ found an application in the description 
of D-branes as solitons of the tachyon field in the limit of large $B$-field \cite{Dasgupta_etal}, 
\cite{Harvey_tach}. This subject was further developed in papers \cite{Witten}, \cite{Sochichiu}, \cite{Strom_D},
\cite{Harvey_D2}, \cite{Harvey_D2}, \cite{K}, \cite{Sen}, \cite{Shenker_etal}, \cite{Li}, \cite{deAlwis}.

Solitons on noncommutative tori and their application to the description of tachyon condensation were studied in 
\cite{Bars_etal}, \cite{Krajewski}, \cite{Matsuo_etal}. Solitonic solutions on other noncommutative spaces such as noncommutative 
orbifolds and fuzzy spheres are considered  in \cite{Hikida_etal}, \cite{orbifolds}, \cite{fuzzy_inst}.

\end{document}